\documentclass[11pt]{article}
\usepackage{amsmath,amssymb,color,epsfig,cite,url}
\usepackage{xcolor}
\usepackage{mathrsfs}

\usepackage{mathtools}

\allowdisplaybreaks

\usepackage{amsfonts}
\def\td{\tilde}
\newcommand{\hoch}[1]{$\, ^{#1}$}

\makeatletter
\@addtoreset{equation}{section}
\makeatother

\newcommand{\be}{\begin{equation}}
\newcommand{\ee}{\end{equation}}
\newcommand{\bea} {\begin{eqnarray}}
\newcommand{\eea}{\end{eqnarray}}
\newcommand{\nn}{\nonumber}

\def\ft#1#2{{\textstyle{\frac{\scriptstyle #1}{\scriptstyle #2} } }}
\def\fft#1#2{{\frac{#1}{#2}}}
\def\dfft#1#2{{\displaystyle\fft{#1}{#2}}}

\def\dfft#1#2{{\displaystyle\fft{#1}{#2}}}

\def\0{{\sst{(0)}}}
\def\1{{\sst{(1)}}}
\def\2{{\sst{(2)}}}
\def\3{{\sst{(3)}}}
\def\4{{\sst{(4)}}}
\def\5{{\sst{(5)}}}
\def\6{{\sst{(6)}}}
\def\7{{\sst{(7)}}}
\def\8{{\sst{(8)}}}
\def\sst#1{{\scriptscriptstyle #1}}

\def\del{{\partial}}

\def\crampest{\medmuskip = 1mu plus 1mu minus 1mu}
\def\uncramp{\medmuskip = 4mu plus 2mu minus 4mu}

\def\wtd{\widetilde}

\def\ie{{ i.e.~}}

\def\ben{\begin{equation}}
\def\bea{\begin{eqnarray}}
\def\een{\end{equation}}
\def\eea{\end{eqnarray}}

\def\ft#1#2{{\textstyle{\frac{\scriptstyle #1}{\scriptstyle #2} } }}
\def\fft#1#2{{\frac{#1}{#2}}}

\begin{document}
\begin{flushright}
\hfill {UPR-1336-T\ \ \ MI-HET-890}\\
\end{flushright}

\begin{center}

{\large {\bf 
Mass and Force Relations for Extremal E2MD Black Holes
}}

\vspace{15pt}
{\large S. Cremonini$^{1}$, M. Cveti\v c$^{2,3,4}$, 
              C.N. Pope$^{5,6}$ and A. Saha$^{5}$}

\vspace{15pt}

{\hoch{1}}{\it Department of Physics, Lehigh University, Bethlehem, 
PA 18018, USA}

{\hoch{2}}{\it Department of Physics and Astronomy,
University of Pennsylvania, \\
Philadelphia, PA 19104, USA}

{\hoch{3}}{\it Department of Mathematics, University of Pennsylvania, Philadelphia, PA 19104, USA}

{\hoch{4}}{\it Center for Applied Mathematics and Theoretical Physics,\\
University of Maribor, SI2000 Maribor, Slovenia}


\hoch{5}{\it George P. \& Cynthia Woods Mitchell  Institute
for Fundamental Physics and Astronomy,\\
Texas A\&M University, College Station, TX 77843, USA}

\hoch{6}{\it DAMTP, Centre for Mathematical Sciences,
 Cambridge University,\\  Wilberforce Road, Cambridge CB3 OWA, UK}


\vspace{10pt}

\begin{abstract}
We study static extremal black holes in Einstein gravity coupled to a dilaton and two Maxwell fields with independent dilaton couplings $a$ and $b$. When $b=-1/a$ in four dimensions, a time-symmetric initial-value construction allows us to determine the masses and interaction energies of multi-black-hole configurations. 
On this special locus, the long-range force between any two extremal black holes vanishes. 
For arbitrary $a$ and $b$, a constant dilaton-shift symmetry, together with homogeneity and the extremality condition, yields a first-order ordinary differential equation that determines the extremal mass as a function of the two electric charges.
This equation allows us to analyze the force between non-identical extremal black holes without having to know
the explicit black-hole geometry. In all analytically controlled regimes that we examine, the locus $b=-1/a$ separates attractive behavior for $b>-1/a$ from repulsive behavior for $b<-1/a$. We extend the analysis to arbitrary spacetime dimensions, where the corresponding force-cancellation condition is $ab=-2(d-2)/(d-1)$. Finally, we test this sign pattern using the exact $A_2$, $B_2$ and $G_2$ Toda black holes. In the $G_2$ case, a potentially problematic branch is excluded because it contains naked singularities outside the horizon, suggesting an intriguing connection between regularity and the sign of long-range forces.
\end{abstract}

\newpage

\end{center}

\tableofcontents

\section{Introduction}

A useful way of probing the constraints that quantum gravity may impose on low-energy effective field theories is to study the relative strength of gravity and the other long-range interactions present in the theory. 
This idea is central to the Weak Gravity Conjecture (WGC) \cite{Arkani-Hamed:2006emk}, which asserts that 
gravity should be the weakest force, in an appropriate sense, in any consistent theory of quantum gravity. 
A closely related statement is the Repulsive Force Conjecture (RFC) \cite{Arkani-Hamed:2006emk,Palti:2017elp,Heidenreich:2019zkl}, which formulates this expectation in theories that include massless scalar fields: a theory coupled to gravity should contain states for which the 
long-range force is repulsive, including gauge, gravitational and scalar exchange.
Indeed, in theories with massless scalar fields, the force between two charged objects is not determined only by their masses and gauge charges. Scalar exchange contributes an additional long-range interaction, whose sign and magnitude depend on the scalar charges of the objects. The relevant comparison is therefore between gravitational attraction, electromagnetic repulsion and scalar-mediated forces. 
Since black holes provide universal charged states in gravitational theories, extremal black holes offer a natural arena in which to test and refine these ideas.

While the RFC is formulated in terms of \emph{identical} extremal black holes, more generally one may ask what happens for \emph{non-identical} extremal black holes carrying different ratios of the available charges. 
Indeed, while for identical black holes the scalar contribution to the 
force is always attractive, this is no longer the case when the solutions are distinct. This is clearly visible from 
the behavior of the 
long range force between two generic solutions, 
\bea
\label{introforce}
{\cal F}_{12} = \frac{Q_1\,Q_2  - M_1\,M_2 - \Sigma_1\,\Sigma_2}{|\vec{x}_{12}|^2}\,,
\eea
where 
$Q_i$, $M_i$ and $\Sigma_i$  ($i=1,2$), denote, respectively, the black hole electric charges\footnote{Here for simplicity we are assuming that the black hole is charged under a single gauge field.}, masses and scalar charges, and $|\vec{x}_{12}|$ denotes the distance between them. 
When the solutions are identical, the scalar contribution to the force $\propto -\Sigma^2$ is clearly attractive, while for generic $\Sigma_i$ it can have either sign. 
Since the gravitational force is always attractive, while the gauge-field contribution is repulsive for positive charges, the scalar fields can play a decisive role in determining the overall sign of the long-range interaction.

In \cite{Cremonini:2022sxf,emdpaper,Cremonini:2024eog} we initiated a program to 
examine  whether long-range forces between distinct states encode useful information about the underlying theory, working with a series 
of models of increasing complexity.
In our earlier work on Einstein-Maxwell-Dilaton theories \cite{emdpaper}, for example, we found that the sign of the force between non-identical extremal black holes is \emph{correlated} with the behavior of the extremal mass as a function of the charges. In particular, we saw that special values of the dilaton coupling can separate regions in which the force is attractive from regions in which it is repulsive, with BPS cases often appearing at the boundary between the two behaviors.
A key step in \cite{emdpaper} was obtaining
a simple first-order ordinary differential equation for the black hole mass in terms of its electric and magnetic charges. In turn, this equation provided a powerful way to derive general results about the long-range interactions between non-identical, extremal black holes.

In this paper we follow a similar procedure to study these questions in a theory containing gravity, a dilaton and two Maxwell fields. In four spacetime dimensions the Lagrangian is
\begin{equation}
\label{introlag}
{\cal L}=\sqrt{-g}\left(R-2(\partial\phi)^2-e^{-2a\phi}F^2-e^{-2b\phi}\widetilde F^2\right),
\end{equation}
with the coupling of the two gauge fields to the dilaton controlled by the parameters $a$ and $b$. This theory provides a simple setting in which to examine extremal black holes carrying two independent electric charges.
It also admits a time-symmetric initial-value formulation when the two dilaton couplings obey a special relation, i.e. $b=-1/a$ in four dimensions and 
$a b = - \frac{2(d-2)}{d-1}$ in $D=d+1$ dimensions.
The initial-value construction gives a direct way to compute the ADM mass, electric and scalar charges and the interaction energy for a collection of black holes.
Closely related use of time-symmetric initial data was made recently in \cite{vinort}, where interactions and equilibrium conditions were studied for multi-black-hole configurations in Einstein-Maxwell-Dilaton theory.
In that work, self-interaction energies were used to distinguish genuinely dynamical initial data from constant-time slices of static solutions, and the resulting interaction energies were shown to reproduce the expected gravitational, electromagnetic and scalar long-range forces.

In the present work, the initial-value analysis provides the starting point for a broader study of extremal black-hole forces in the theory (\ref{introlag}). Although the initial-data construction applies directly only on the special locus relating $a$ and $b$, we show that the extremal mass can be constrained much more generally, for  arbitrary values of the dilaton couplings. The key observation is that the theory has a constant dilaton-shift symmetry, accompanied by suitable rescalings of the two Maxwell fields. This symmetry fixes the scaling behavior of the electric charges and relates the scalar charge to derivatives of the extremal mass with respect to the charges. Combining this relation with the extremality condition yields a first-order ordinary differential equation for a single function $h(x)$ that encodes the extremal mass, electric charges and scalar charge.

This mass equation provides a practical way to diagnose the long-range interaction between non-identical extremal black holes without having to know the full spacetime solution. On the special locus $b=-1/a$, the extremal mass is linear in the two charges and the force between \emph{any} two extremal black holes (i.e., for arbitrary charge ratios) vanishes.
In every analytically controlled regime that we examine -- perturbatively near this locus, near the point of vanishing scalar charge, and in asymptotic charge-ratio limits -- we find the same sign pattern for the long-range force: 
\begin{equation}
\label{forcecondition}
b>-{1\over a} \;\; \Longrightarrow \;\; \text{attractive}\, ,
\qquad
b<-{1\over a} \;\; \Longrightarrow \;\; \text{repulsive} \, .
\end{equation}
Thus, the special locus $b=-1/a$ acts as a BPS-like condition, i.e. a dividing line between qualitatively different force behaviors and on which the mutual force between two non-equal balck holes vanishes.
The exact $B_2$ and $G_2$ Toda black holes we examine in Appendix B provide nontrivial tests of this behavior. In the $G_2$ case, a particular branch which superficially appears to evade the pattern (\ref{forcecondition}) is in fact excluded by the presence of naked singularities outside the horizon. 
This reveals an interesting interplay between regularity and the allowed behavior of the long-range force. We also extend the aforementioned coupling-dependent feature of the mutual force to an arbitrary higher dimension $D=d+1$, where we show that the special locus which shows BPS-like condition is governed by $b = -\frac{2(d-2)}{(d-1)\,a}$, and the force is attractive or repulsive depending on whether $b$ is respectively greater or less than this value.

The outline of the paper is as follows. In section \ref{initvalsec}  we review the four-dimensional time-symmetric initial-value construction for $b=-1/a$ and use it to derive the interaction energy and long-range force between two black holes.  In section \ref{massform4} we derive a mass equation for extremal black holes with arbitrary dilaton couplings $a$ and $b$, and discuss exact and approximate solutions. 
In section \ref{sec4} we apply this equation to the force between non-identical extremal black holes, showing that the locus $b=-1/a$ separates attractive from repulsive behavior in several regimes. In section \ref{sec5} we 
present explicit black-hole solutions on this special locus. 
Section \ref{sec6} extends the analysis to higher dimensions, and Appendix~A derives the extremality condition for arbitrary couplings.
Appendix~B examines the exceptional $A_2, B_2$ and $G_2$ 
Toda black holes, derives their mass functions, and analyzes the allowed parameter branches and the corresponding forces.

\section{\label{initvalsec} 
   Initial Data and Interaction Energy}

\subsection{The initial data}

  In this section we use the results in section 3.4 of \cite{cvegibpop}, 
with a minor change of notation.
The system we are interested in comprises of gravity with a dilaton and two Maxwell fields, denoted
by $F$ and $\wtd F$.  The Lagrangian is
\bea
{\cal L}= \sqrt{-g}\, \Big(R -2 (\del\phi)^2 - e^{-2a\phi}\, F^2  -
  e^{-2b\phi}\, \wtd F^2\Big)\,.\label{2fieldlag}
\eea
In order to be able to solve for time-symmetric initial data, we
must take
\bea
b = -\fft1{a}\,,\label{barel}
\eea
with $a$ allowed to be arbitrary.

  As shown in \cite{cvegibpop}, we can take the following ansatz for 
time-symmetric initial data:
\bea
\Phi^4&=& (C\, D)^{\ft{2}{1+a^2}}\, (\wtd C\, \wtd D)^{\ft{2 a^2}{1+a^2}}\,,
\qquad e^\phi = \Big(\fft{C\, D}{\wtd C\, \wtd D}\Big)^{-\ft{a}{1+a^2}}\,,
\nn\\
E_\alpha&=& \fft1{\sqrt{1+a^2}}\, 
\Big(\fft{C\, D}{\wtd C\, \wtd D}\Big)^{-\ft{a^2}{1+a^2}}\,\del_\alpha
\log\Big(\fft{C}{D}\Big)\,,\nn\\
\wtd E_\alpha&=& \fft{a}{\sqrt{1+a^2}}\,
\Big(\fft{C\, D}{\wtd C\, \wtd D}\Big)^{\ft{1}{1+a^2}}\,\del_\alpha
\log\Big(\fft{\wtd C}{\wtd D}\Big)\,,\label{initial}
\eea
with the 3-metric being given by $ds_3^2 = \Phi^4\, dx^\alpha\,dx^\alpha$,
and $C, D, \wtd C$ and $\wtd D$ being arbitrary harmonic functions in
the Euclidean 3-metric $dx^\alpha\,dx^\alpha$.

   The four harmonic functions may be taken to have the ${\cal N}$-centre form
\bea
C&=& 1 + \sum_{i=1}^{\cal N} \fft{c_i}{|\vec x- \vec x_i|}\,,\qquad
D= 1+\sum_{i=1}^{\cal N} \fft{d_i}{|\vec x- \vec x_i|}\,,\nn\\
\wtd C&=& 1 + \sum_{i=1}^{\cal N} \fft{\td c_i}{|\vec x- \vec x_i|}\,,\qquad
\wtd D= 1+\sum_{i=1}^{\cal N} \fft{\td d_i}{|\vec x- \vec x_i|}\,,
\eea
where $c_i, d_i, \td c_i$ and $\td d_i$ are arbitrary constants.  These
time-symmetric initial data describe a system of
${\cal N}$
black holes, located
at the points $\vec x_i$.

  The total mass of the system of ${\cal N}$ black holes can be found by looking
at the asymptotic form of the 3-metric $ds_3^2$, at distances such
that $|\vec x| >> |\vec x_i|$ for all $i$.  Thus we have 
\bea
\Phi^4 = 1 + \fft{2}{1+a^2}\, \sum_{i=1}^{\cal N} \Big[c_i+d_i + 
           a^2\, (\td c_i + \td d_i)\Big]\, \fft1{|\vec x|} + \cdots
\eea
and since this will have the form $\Phi^4=1 + \dfft{2M}{|\vec x|} +\cdots$
we can read off the total mass as
\bea
M= \fft{1}{1+a^2}\, \sum_{i=1}^{\cal N} \Big[c_i+d_i +
           a^2\, (\td c_i + \td d_i)\Big] \,.\label{totalmass}
\eea

   The total electric charges $Q$ and $\wtd Q$ 
can be found by performing the Gaussian integrals
\bea
Q=\fft1{4\pi}\,\int d\Sigma^\alpha\, \Phi^2 e^{-2a\phi}\, E_\alpha\,,\qquad
\wtd Q=\fft1{4\pi}\,\int d\Sigma^\alpha\, \Phi^2 e^{-2b\phi}\, \wtd E_\alpha\,.\label{totalcharges}
\eea
From the form of the initial data for the electric fields in eqns 
(\ref{initial}), it follows that the electric charges will be given by
\bea
Q= \fft1{\sqrt{1+a^2}}\, \sum_{i=1}^{\cal N} (c_i-d_i)\,,\qquad
\wtd Q= \fft{a}{\sqrt{1+a^2}}\, \sum_{i=1}^{\cal N} (\td c_i-\td d_i)\,.
\label{totalcharges2}
\eea

   A total scalar charge $\Sigma$ for the dilaton field can be defined as
the coefficient of $\dfft{1}{|\vec x|}$ in the large-$|x|$ expansion of
the field $\phi$.  From the expression for $\phi$ in eqns (\ref{initial}),
it follows that
\bea
\Sigma = -\fft{a}{1+a^2}\, \sum_{i=1}^{\cal N} (c_i+d_i-\td c_i -\td d_i)\,.
\eea

\subsection{Interactions}

  Following discussions by Brill and Lindquist \cite{brillind}, one can 
investigate questions such as the interaction energy between the 
black holes.  In this part of the discussion, we shall rather closely 
follow the procedures described in \cite{vinort}, especially, section 4.6
of that paper.\footnote{In fact the discussion in section 4.6 of \cite{vinort}
is equivalent to a special case of our discussion here, in which the dilaton
couplings are $a=1$ and $b=-1$.  In this special case, our two-field
EMD theory for two electric charges is equivalent, for the present purposes, 
to a single-field EMD theory with $a=1$ and with electric and magnetic 
charges.}

 Consider first the case of a single black hole, \ie the case $\mathcal{N}=1$, and
where furthermore the electric charges and the dilaton field are set to
zero; this is the case where $C, D, \wtd C$ and $\wtd D$ are all set equal,
and the non-zero constants obey $c_1=d_1=\td c_1=\td d_1$, 
which we can just call $c$.
In this case 
%
%
we
have precisely the initial data for the Schwarzschild
black hole.  Without loss of generality we can take $\vec x_1=0$, so the
black hole is located at the origin.  The 3-metric can then be written using 
spherical polar coordinates, 
 with $\vec x=(\rho\, \sin\theta\,\cos\varphi,
\rho\, \sin\theta\, \sin\varphi,\rho \cos\theta)$, and it takes the form
\bea
ds_3^2= \Phi^4\, dx^\alpha\,dx^\alpha= 
\Big(1+\fft{c}{\rho}\Big)^4\,(d\rho^2 + \rho^2\, d\Omega^2)\,.
\label{rhomet}
\eea
There is an exact inversion symmetry, where we define
\bea
\rho'= \fft{c^2}{\rho}\,,
\eea
for which the metric (\ref{rhomet}) becomes
\bea
ds_3^2= 
\Big(1+\fft{c}{\rho'}\Big)^4\,(d{\rho'}^2 + {\rho'}^2\, d\Omega^2)\,.
\label{rhopmet}
\eea
Thus one can pass from an asymptotic region near $\rho=\infty$ to
another exactly equivalent asymptotic region near $\rho=0$ (\ie near
$\rho'=\infty$).  This is the familiar Schwarzschild picture of two
asymptotic regions connected by an Einstein-Rosen bridge.  It follows
from the equivalence of the two descriptions of the metric that the
mass, as measured in the asymptotic region near $\rho=0$, will be
equal to the mass as measured in the asymptotic region near
$\rho=\infty$, that is, $M=2 c$.  

The utility of the above observation is that, when we consider $\mathcal{N}$ black holes, we can associate a mass
$M_i$  specifically to the $i$'th black hole by taking the
full $\mathcal{N}$-hole metric, focusing on the region where $\vec x$ approaches
$\vec x_i$, and then inverting through the point $\vec x_i$ in order
to obtain an asymptotic region dominated entirely by the effects of
the $i$'th black hole.
In a similar vein, when the charges and dilaton are also turned on, 
we can use this inversion procedure to define electric charges $Q_i$ and $\wtd Q_i$ 
and a scalar charge $\Sigma_i$, associated specifically with
the $i$'th black hole.

When $\vec x$ is close to the location $\vec x_i$ of a specific
black hole, we shall have
\bea
C &\approx& 1 + \fft{c_i}{|\vec x-\vec x_i|} + A_i\,,\qquad
   A_i = \sum_{j\ne i} \fft{c_j}{|\vec x_{ij}|}\,,\nn\\
D &\approx& 1 + \fft{d_i}{|\vec x-\vec x_i|} + B_i\,,\qquad
   B_i = \sum_{j\ne i} \fft{d_j}{|\vec x_{ij}|}\,,\nn\\
\wtd C &\approx& 1 + \fft{\td c_i}{|\vec x-\vec x_i|} + \wtd A_i\,,\qquad
   \wtd A_i = \sum_{j\ne i} \fft{\td c_j}{|\vec x_{ij}|}\,,\nn\\
\wtd D &\approx& 1 + \fft{\td d_i}{|\vec x-\vec x_i|} + \wtd B_i\,,\qquad
   \wtd B_i = \sum_{j\ne i} \fft{\td d_j}{|\vec x_{ij}|}\,,\label{ABdefs}
\eea
where $\vec x_{ij}= \vec x_i - \vec x_j$.
Using now spherical polar coordinates defined by
\bea
\vec x - \vec x_i = (\rho\, \sin\theta\,\cos\varphi,
\rho\, \sin\theta\, \sin\varphi, \rho \cos\theta)\,,
\eea
we can perform an inversion by defining
\bea
\rho' = \fft{\alpha_i^2}{\rho}\,,\qquad
\alpha_i^2 = \big(c_i\, d_i\big)^{\ft1{1+a^2}}\, 
           \big(\td c_i\, \td d_i\big)^{\ft{a^2}{1+a^2}}\,,\label{invalpha}
\eea
which will map the 3-metric into the form
\bea
ds_3^2 \approx {\Phi'}^4\, (d{\rho'}^2 + {\rho'}^2\, d\Omega^2)\,,
\eea
with
\bea
{\Phi'}^4 &=& \Big[ \Big(1 + \fft{\alpha_i^2\, (1+A_i)}{c_i\, \rho'}\Big)
 \Big(1 + \fft{\alpha_i^2\, (1+B_i)}{d_i\, \rho'}\Big)\Big]^{\ft{2}{1+a^2}}
\times\nn\\
&&
 \Big[ \Big(1 + \fft{\alpha_i^2\, (1+\wtd A_i)}{\td c_i\, \rho'}\Big)
 \Big(1 + \fft{\alpha_i^2\, (1+\wtd B_i)}{\td d_i\, \rho'}\Big)
   \Big]^{\ft{2a^2}{1+a^2}}\,.
\eea
Thus in the asymptotic region where $\rho'$ goes to infinity, we can read off
the mass
\bea
M_i = \fft{\alpha_i^2}{1+a^2}\,
  \Big[\fft{1+A_i}{c_i} + \fft{1+B_i}{d_i} + 
   \fft{a^2\, (1+\wtd A_i)}{\td c_i} + \fft{a^2\, (1+\wtd B_i)}{\td d_i}
\Big]\,.\label{massMi}
\eea

An important constraint to keep in mind is that, for the case of a single
black hole, 
%
there should be no ``self-energy''
corresponding to a discrepancy between the original mass $M$ given 
in eqn (\ref{totalmass}) (for ${\cal N}=1$) and the expression $M_i$ in
eqn (\ref{massMi}), with $i=1$. 
In order for the two masses
to be equal we must require the integration constants to be such that
$c_1\, d_1 = \td c_1\, \td d_1$.  
Of course, we will also require
such conditions to hold for each black hole center 
in the general ${\cal N}$ case. Thus, we 
impose
\bea
   c_i\, d_i = \td c_i\, \td d_i\,,\qquad \hbox{for all $i$}\,.
\label{noselfenergy}
\eea
It then follows from the expression for $\alpha_i^2$ in eqn (\ref{invalpha})
that for all $i$ we will have
\bea
\alpha_i^2 = c_i\, d_i = \td c_i\, \td d_i\,.\label{alphacd}
\eea
Assuming these conditions are imposed, it then follows from
eqn (\ref{massMi}) that
\bea
M_i = \fft1{1+a^2}\, \Big[(1+A_i)\, d_i + (1+B_i)\, c_i +
                    a^2\, (1+\wtd A_i)\, \td d_i + a^2\, (1+\wtd B_i)\, \td c_i
\Big]\,.\label{finalmassMi}
\eea

The electric charges as measured in the asymptotic region $\rho'\longrightarrow\infty$
obtained by the inversion through the $i$'th black hole are given by
\bea
Q_i &=& \fft{\alpha_i^2}{\sqrt{1+a^2}}\, \Big[\fft{1+A_i}{c_i} 
                   -\fft{1+B_i}{d_i}\Big]\,\nn\\
&=& \fft1{\sqrt{1+a^2}}\, \big[(1+A_i)\, d_i - (1+B_i)\, c_i\big]\,,\label{finalQi}
\eea
and
\bea
\wtd Q_i &=& \fft{a\, \alpha_i^2}{\sqrt{1+a^2}}\, \Big[\fft{1+\wtd A_i}{\td c_i}
                   -\fft{1+\wtd B_i}{\td d_i}\Big]\, \nn\\
&=& \fft{a}{\sqrt{1+a^2}}\, \big[(1+\wtd A_i)\, \td d_i - 
              (1+\wtd B_i)\, \td c_i\big]\,,\label{finalQQi}
\eea
where in each case, the second line follows upon using the conditions 
(\ref{alphacd}).  

The scalar charge $\Sigma_i$ in the inverted asymptotic region of
the $i$'th black hole can be read off as the coefficient of $\dfft1{\rho'}$
in the expansion for $\phi$.  This gives
\bea
\Sigma_i &=& -\fft{a\,\alpha_i^2}{1+a^2}\, \Big[
 \fft{1+A_i}{c_i} + \fft{1+B_i}{d_i} - \fft{1+\wtd A_i}{\td c_i} -
     \fft{1+\wtd B_i}{\td d_i}\Big]\,\nn\\
&=& -\fft{a}{1+a^2}\,\big[(1+A_i)\, d_i + (1+B_i)\, c_i -
   (1+\wtd A_i)\, \td d_i - (1+\wtd B_i)\, \td c_i\big]\,,\label{finalSigi}
\eea
where again, the second line follows upon using the conditions (\ref{alphacd}).

Finally, the interaction energy for the system of $\mathcal{N}$ black holes can be defined to
be the total mass $M$ given in eqn (\ref{totalmass}) minus the sum of the
individual masses $M_i$ given in eqn (\ref{finalmassMi}).  Thus,
\bea
M_{\rm int} &=& M -\sum_i M_i\nn\nn\\
&=& M- \fft1{1+a^2}\, \sum_i  \Big[(1+A_i)\, d_i + (1+B_i)\, c_i +
                    a^2\, (1+\wtd A_i)\, \td d_i + a^2 \, (1+\wtd B_i)\, \td c_i
\Big]\nn\\
&=& -\fft1{1+a^2}\, \sum_i  \Big[A_i\, d_i + B_i\, c_i +
                    a^2\, \wtd A_i\, \td d_i + a^2\, \wtd B_i\, \td c_i\Big]\,.
\eea
 From the definitions of $A_i, B_i, \wtd A_i$ and $\wtd B_i$ in eqns 
(\ref{ABdefs}), it follows that the interaction energy is given by
\bea
M_{\rm int} = -\fft2{1+a^2}\, 
\sum_i\sum_{j\ne i}\fft{c_i\, d_j + a^2\, \td c_i\, \td d_j}{|\vec x_{ij}|}\,.
\label{Mintres}
\eea

\subsection{Interaction energy and force between two black holes \label{Intenergy4D}}

  If we consider specifically the case of two black holes, \ie $\mathcal{N}=2$,
we can obtain simple explicit 
expressions, to leading order in $\dfft1{|\vec x_{12}|}$ (the inverse
of the separation distance between the two black holes), 
for the eight integration
constants $(c_1, c_2, d_1, d_2, \td c_1, \td c_2, \td d_1,\td d_2)$ in terms
of the eight black hole parameters $(M_1, M_2, Q_1, Q_2, \wtd Q_1, \wtd Q_2,
\Sigma_1, \Sigma_2)$.  Using these, we can then obtain a
simple relation, at leading order, for the interaction energy between the
two black holes, expressed in terms of the black hole parameters.

   Taking ${\cal N}=2$, it can be seen from the expressions given above that, working 
at leading order (\ie neglecting all terms of order $|\vec x_{12}|^{-1}$ in the right-hand sides of
eqns (\ref{finalmassMi}), (\ref{finalQi}), (\ref{finalQQi}) and (\ref{finalSigi})), one has
\bea
M_1&=& \fft{c_1+d_1 + a^2\,(\td c_1+\td d_1)}{1+a^2}\,,\qquad
M_2= \fft{c_2+d_2 + a^2\,(\td c_2+\td d_2)}{1+a^2}\,,\nn\\
Q_1 &=& \fft{d_1-c_1}{\sqrt{1+a^2}}\,,\qquad
Q_2=\fft{d_2-c_2}{\sqrt{1+a^2}}\,,\nn\\
\wtd Q_1 &=& \fft{a\,(\td d_1-\td c_1)}{\sqrt{1+a^2}}\,,\qquad
\wtd Q_2=\fft{a\,(\td d_2-\td c_2)}{\sqrt{1+a^2}}\,,\nn\\
\Sigma_1 &=& -\fft{a\, (c_1+d_1-\td c_1-\td d_1)}{1+a^2}\,,\qquad
\Sigma_2 = -\fft{a\, (c_2+d_2-\td c_2-\td d_2)}{1+a^2}\,.
\eea
These equations can be solved for the constants $(c_i,d_i,\td c_i, \td d_i)$,
giving
\bea
c_1&=& \fft12 \big[ M_1-\sqrt{1+a^2}\, Q_1- a\, \Sigma_1\big]\,,\qquad
c_2= \fft12 \big[ M_2-\sqrt{1+a^2}\, Q_2- a\, \Sigma_2\big]\,,\nn\\
d_1&=& \fft12 \big[ M_1+\sqrt{1+a^2}\, Q_1- a\, \Sigma_1\big]\,,\qquad
d_2= \fft12 \big[ M_2+\sqrt{1+a^2}\, Q_2- a\, \Sigma_2\big]\,,\nn\\
\td c_1&=& \fft1{2a} 
  \big[a\, M_1-\sqrt{1+a^2}\, \wtd Q_1 + \Sigma_1\big]\,,\quad
\td c_2= \fft1{2a} 
\big[a\, M_2-\sqrt{1+a^2}\, \wtd Q_2+  \Sigma_2\big]\,,\nn\\
\td d_1&=& \fft1{2a} 
  \big[a\, M_1+\sqrt{1+a^2}\, \wtd Q_1+  \Sigma_1\big]\,,\quad
\td d_2= \fft1{2a} 
    \big[a\, M_2+\sqrt{1+a^2}\, \wtd Q_2 + \Sigma_2\big]\,.\label{cdsols}
\eea
Plugging these into the expression for $M_{\rm int}$ given in eqn 
(\ref{Mintres}), specialised to the $N=2$ case, gives the interaction energy
between the two black holes as
\bea
M_{\rm int}= \frac{-\big( M_1\, M_2 - Q_1\, Q_2 - \wtd Q_1\, \wtd Q_2 +
            \Sigma_1\, \Sigma_2\big)}{| \vec{x}_{12}|}\,.\label{Mintexp}
\eea
 
   From the solutions (\ref{cdsols}), it can be seen that the condition
(\ref{noselfenergy}) which ensures that each black hole has no self-energy implies
\bea
(1-a^2)\, {\Sigma}^2 + 2 a \, M\, \Sigma + a^2\, Q^2 - \wtd Q^2=0\,,
\label{Sigmacond}
\eea
expressed in terms of the charges of the particular black hole under consideration.
In other words, the condition (\ref{noselfenergy}) implies that the
scalar charge must be expressible in terms of the mass and the electric 
charges.

  Consider now the interaction energy (\ref{Mintexp}) between two 
black holes, and specialise to the case where they
are identical, so $M_1=M_2=M$, \ $Q_1=Q_2=Q$,\ $\wtd Q_1=\wtd Q_2=\wtd Q$
and $\Sigma_1=\Sigma_2$:
\bea
 M_{\rm int} = \frac{-M^2 + Q^2 + \wtd Q^2 -\Sigma^2}{|\vec{x}_{12}|}\,.\label{identMint}
\eea
If the identical 
black holes are furthermore extremal, then
the interaction energy between them will be zero, so\footnote{Here, we have obtained the zero-force equation (\ref{extremalcon}) from the initial-value formulation, which applies specifically for the cases where the dilaton couplings $a$ and $b$ are related by $b=-\dfft1{a}$.
In appendix \ref{D4con} we show that the same formula (\ref{extremalcon}) holds also in the general case where the dilaton couplings $a$ and $b$ are arbitrary.}
\bea
M^2 - Q^2 - \wtd Q^2 + \Sigma^2=0\,.\label{extremalcon}
\eea
Eliminating the scalar charge between this equation and eqn (\ref{Sigmacond})
then implies that
\bea
\Big[M-\fft{Q+a\wtd Q}{\sqrt{1+a^2}}\Big]\,
 \Big[M-\fft{Q-a\wtd Q}{\sqrt{1+a^2}}\Big]\,
 \Big[M+\fft{Q+a\wtd Q}{\sqrt{1+a^2}}\Big]\,
 \Big[M+\fft{Q-a\wtd Q}{\sqrt{1+a^2}}\Big]=0\,.
\eea
For given $Q$ and $\wtd Q$, 
two of these factors vanish for positive $M$ and the
other two vanish for negative $M$.  Suppose, for example, we have
\bea
Q>0\,,\qquad \hbox{and}\qquad  Q>|a\, \wtd Q|\,.
\eea
Then, the two roots with positive $M$ will be 
\bea
M= \fft{Q+a\, \wtd Q}{\sqrt{1+a^2}}\,,\qquad \hbox{and}\qquad
M= \fft{Q-a\, \wtd Q}{\sqrt{1+a^2}}\,.\label{BPSconds}
\eea
In either of these circumstances, the black holes obey a BPS condition.

   The BPS conditions in eqns (\ref{BPSconds}) can be substituted back
into the extremality (\ref{extremalcon}).  Concretely, let us take
\bea
M= \fft{Q+a\, \wtd Q}{\sqrt{1+a^2}}\,.\label{solM11}
\eea
Up to the sign of $\Sigma$ 
and the sign of $a$ (the latter as in eqns (\ref{BPSconds})), we can then solve
for $\Sigma$ to find 
\bea
\Sigma= \fft{\wtd Q- a\, Q}{\sqrt{1+a^2}}\,.\label{solSig}
\eea
One can now use eqns (\ref{solM11}) and (\ref{solSig}) for two
black holes labelled by 1 and 2, with charges $(Q_1,\wtd Q_1)$ and 
$(Q_2,\wtd Q_2)$, and hence calculate the interaction energy
in eqn (\ref{Mintexp}) for two {\it non-identical} extremal black holes.
The interaction energy turns out always to be zero.

\section{A Mass Formula for Extremal Black Holes\label{massform4}}

  The Lagrangian (\ref{2fieldlag}) has a global symmetry under a constant
dilaton shift, combined with appropriate rescalings of the Maxwell 
potentials and field strengths:
\bea
\phi\longrightarrow \phi+\phi_0\,,\qquad
 F\longrightarrow e^{a\,\phi_0}\, F\,,\qquad
\wtd F\longrightarrow e^{b\, \phi_0}\, \wtd F\,.\label{globalsym}
\eea
It follows that the conserved electric charges
\bea
Q=\fft1{4\pi}\, \int e^{-2a\,\phi}\, {*F}\,,\qquad 
\wtd Q=\fft1{4\pi}\, \int e^{-2b\,\phi}\, {*\wtd F}
\eea
will scale as 
\bea
Q\longrightarrow e^{-a\,\phi_0}\, Q\,,\qquad
\wtd Q\longrightarrow e^{-b\,\phi_0}\, \wtd Q\,.\label{chargescaling}
\eea
   It should be noted that the global symmetry described above is valid
for {\it all} choices of the constants $a$ and $b$.  Thus, in what follows
we shall not impose the relation (\ref{barel}); rather, we shall be
taking $a$ and $b$ to be {\it two independent constants}.

  For extremal black hole solutions, the mass will be a function only of
$Q$ and $\wtd Q$, so we may write $M= M(Q,\wtd Q)$.  
Since $M$, $Q$ and $\wtd Q$ all have the same dimensions, it must be that
$M(Q,\wtd Q)$ obeys the homogeneous scaling relation
\bea
M(\lambda\, Q,\lambda\, \wtd Q)= \lambda\, M(Q,\wtd Q)\,.\label{masshomo}
\eea
Differentiating this with respect to $\lambda$ and then setting 
$\lambda=1$ implies
\bea
M = Q\, \fft{\del M}{\del Q} + \wtd Q\, \fft{\del M}{\del \wtd Q}\,.
\label{massdiff}
\eea

As we already mentioned in the previous section,    the scalar charge $\Sigma$ 
is defined to be the coefficient of $\dfft1{r}$
in a large-$r$ expansion of the dilaton field $\phi$.  As shown in 
\cite{gibkalkol}, if the asymptotic value of the dilaton field is 
taken to be $\phi_0$, then the scalar charge can be written as
\bea
\Sigma = \fft{\del M}{\del\phi_0}\,.
\eea
By the chain rule, this can be rewritten as 
\bea
\Sigma= \fft{\del M}{\del Q}\, \fft{\del Q}{\del\phi_0} +
    \fft{\del M}{\del \wtd Q}\, \fft{\del\wtd Q}{\del\phi_0}\,.
\eea
Since we can introduce an asymptotic value for the dilaton field by
using the global symmetry (\ref{globalsym}), which implies
the charge scalings (\ref{chargescaling}), it follows that, after
setting $\phi_0$ to zero, we shall have
\bea
\fft{\del Q}{\del\phi_0} = -a\, Q\,,\qquad 
\fft{\del \wtd Q}{\del\phi_0} = -b\, \wtd Q\,,\label{dhargedphi0}
\eea
and thus,
\bea
\Sigma= -a\, Q\, \fft{\del M}{\del Q} - 
            b\, \wtd Q\, \fft{\del M}{\del \wtd Q}\,.\label{Sigmaform}
\eea

   Since an extremal black hole must satisfy the no-force condition\footnote{Although eqn (\ref{extremalcon}) was derived using the initial-value construction with
$b=-1/a$, Appendix A shows that the same extremality/no-force condition
holds for arbitrary $a$ and $b$.}
given in eqn (\ref{extremalcon}), it 
follows from eqn (\ref{Sigmaform}) that its parameters will obey the
equation
\bea
M^2 -Q^2 -\wtd Q^2 + \Big(a\, Q\, \fft{\del M}{\del Q} +
   b\, \wtd Q\, \fft{\del M}{\del \wtd Q}\Big)^2=0\,,\label{diffeq0}
\eea
where we recall that $M$ is $M(Q,\wtd Q)$.
Using eqn (\ref{massdiff}), eqn (\ref{diffeq0}) can be rewritten as
\bea
M^2-Q^2-\wtd Q^2 +\Big[ (a-b)\, Q\, \fft{\del M}{\del Q} + b M\Big]^2=0\,.
\label{diffeq1}
\eea
This is a differential equation involving only $Q$ derivatives, for which
we can treat $\wtd Q$ simply as a parameter.  It is convenient now to
change the independent variable from $Q$ to $x$, which is defined by
\bea
\label{xdef}
Q=\wtd Q\, e^{2\lambda\, x}\,,
\eea
where $\lambda$ is a constant that will be chosen later.  We may also choose
to write $M(Q,\wtd Q)$ as
\bea
M= \sqrt{Q\, \wtd Q}\, f(x)\,.
\eea
The no-force condition given by eqn (\ref{diffeq1}) then becomes
\bea
f^2 - e^{2\lambda\, x} - e^{-2\lambda\, x} + 
 \Big[\ft12 (a+b)\, f +\fft{a-b}{2\lambda}\, f'\Big]^2=0\,,\label{diffeq2}
\eea
where the prime denotes a derivative with respect to $x$.
The equation can be simplified by
introducing a new variable $h(x)$ and 
choosing the constant $\lambda$ as 
follows\footnote{We have assumed $a\neq b$, so that the variable $x$
is a non-degenerate parametrization of the charge ratio. The case $a=b$,
in which the two Maxwell fields have the same dilaton coupling, is degenerate
from this point of view and should be treated separately, or obtained as a
smooth limit.}:
\bea
h(x) = e^{\mu\, x}\, f(x)\,,\qquad \mu= \ft12 (a+b)\,,\qquad
\lambda= \ft12 (a-b)\,.
\eea
Eqn (\ref{diffeq2}) then becomes
\bea
{h'}^2 + h^2 = e^{2a\, x} + e^{2b\, x}\,.\label{diffeq}
\eea
 From the definitions above, the mass, the electric charge $Q$ and the 
scalar charge
are given in terms of $x$ and $\wtd Q$ by
\bea
M= \wtd Q\, e^{-b\, x}\, h(x)\,,\qquad Q= \wtd Q\, e^{(a-b)\, x}\,,\qquad
\Sigma=  -\wtd Q\, e^{-b\, x}\, h'(x)\,.\label{MSigsols}
\eea
Thus $M(Q,\wtd Q)$ and $\Sigma(Q,\wtd Q)$ for extremal black holes 
are fully determined in terms of $Q$ 
and $\wtd Q$, once the differential equation 
(\ref{diffeq}) is solved for $h(x)$.  These results hold for the entire
2-parameter class of theories described by the Lagrangian (\ref{2fieldlag}),
with $a$ and $b$ being arbitrary constants.

\subsection{Exact solution for $h(x)$ when $b=-\dfft1{a}$}

  It is not obvious in general how to solve the differential equation 
(\ref{diffeq}) explicitly.  However, we can easily find the desired solution
in the case of the extremal black holes of the sub-class of theories where
$b=-\dfft1{a}$, since we studied these in section \ref{initvalsec} when
we discussed the initial-value formulation.  Up to essentially unimportant
sign choices, we arrived at the conclusion there (see eqns (\ref{BPSconds})) that an extremal black hole in 
this class of theories will obey the BPS condition
\bea
M= \fft{Q + a\, \wtd Q}{\sqrt{1+a^2}}\,.\label{extremalmass}
\eea
Plugging the expressions for $M$ and $Q$ given in eqns (\ref{MSigsols}) into
this, we can solve for $h(x)$, finding
\bea
h(x) = \fft1{\sqrt{1+a^2}}\, \big(e^{a\, x} + a\, e^{b\, x}\big)\,. \label{hsolBPS}
\eea
It is straightforward to verify that this does indeed satisfy eqn 
(\ref{diffeq}), provided that the condition $b=-\dfft1{a}$ holds.

In appendix \ref{Todasec} we shall discuss three explictly-known solutions that do not fall into the general class of $b=-\dfft1a$ solutions described above.  These further cases are more complicated, and arise only for very specific values of $a$ and $b$.  They are associated with $A_2$, $B_2$ and $G_2$ Toda systems.  

\subsection{Approximate solution for $h(x)$}

\subsubsection{Perturbing around $b=-\dfft1a$}

We may try to solve for $h(x)$ perturbatively around $b=-\dfft{1}{a}$ by writing $b=-\dfft{1}{a}+\epsilon$, where $\epsilon \in \mathbb{R}$. Such a solution up to order $\epsilon$ is given by
\bea
h(x) &=& h_0(x) + \epsilon\, h_1(x)\,,\nn\\
h_0(x) &=& \frac{1}{\sqrt{1+a^2}}\,(e^{a x} + a\,e^{(-\ft1a + \epsilon)\, x})\,,  \nn\\
h_1(x) &=& \frac{a^2\,\Big(e^{-\frac{x}{a}} + e^{a x}\,(c_1 + (1+a^2)\,x) \Big)}{(1+a^2)^{\fft32}\,\left(1-a\,e^{\ft{(1+a^2)\,x}{a}} \right)}\,, \label{hsolpert1}
\eea
where $c_1$ is an integration constant. The integration constant $c_1$ can be determined by imposing an appropriate boundary condition on $h(x)$.  Define the special point $x=x_0$ such that
\bea
h'(x_0)  = 0.\label{hpbc}
\eea
By differentiating \eqref{diffeq} to give $2 h'\,h'' + 2 h\, h' =2a\, e^{2ax} + 2b\, e^{2bx}$, we see that in order for the condition (\ref{hpbc}) to hold we must choose
\bea
x_0 = \frac{1}{2\,(a-b)}\,\log\left(-\frac{b}{a}\right)\,.
\eea
By virtue of the last equation in \eqref{MSigsols}, the scalar charge 
$\Sigma$ also vanishes at $x=x_0$. The leading (i.e. $\epsilon^0$) order term in $h(x)$ above satisfies this initial condition. At order $\epsilon$ we can use this initial condition to determine the integration constant $c_1$. Writing $b$ as $b=-1/a+\epsilon$ and expanding up to order $\epsilon$ we find
\bea
x_0 &=& -\frac{a\,\log a}{1+a^2} - \frac{a^2\,(1+a^2+2\,\log a)}{2\,(1+a^2)^2}\,\epsilon\,, \nn\\
h'(x)\vert_{x_0} &=& 0 \quad \implies \quad c_1 = a\,(\log a -1)\,,
\eea
which is the choice for $c_1$ that we shall make from now on.

\subsubsection{Perturbing around the point of vanishing scalar charge, $x = x_0$}

Another way to solve the mass ODE perturbatively is to find a series solution for $h(x)$ around $x_0$ for arbitrary values of $b,\, a$. We choose $x_0$ because $h'(x)$ vanishes at that point (c.f. \eqref{hpbc}) 
and we write $h(x)$ as:
\bea
h(x) &=& \bar h_0 + \bar h_2\,(x-x_0)^2 + \bar h_3\,(x-x_0)^3 + \ldots\,, \nn\\
\bar h_0 &=& \sqrt{ 1 - \fft{b}{a}}\,e^{b\,x_0}\,, \nn\\
\bar h_2 &=& -\fft14\, \sqrt{1-\fft{b}{a}}\, (1-\sqrt{1-8\,a\,b})\,e^{b\,x_0}\,, \nn\\
\bar h_3 &=& - \fft{\sqrt{a}\,b\,\sqrt{a-b}\,(a+b)\,(1+3\,\sqrt{1-8\,a\,b})\,e^{b\,x_0}}{6\,(1-9\,a\,b)}\,. \label{hpertx0}
\eea
Since $\Sigma\propto h^\prime(x) \propto (x-x_0)$, this case corresponds to expanding around the point of vanishing scalar charge.
Note that without loss of generality we can assume $a>0$, which implies that $b<0$ in order for $x_0$ to remain real. In particular this means the square roots appearing in the expressions above will always stay real. Also note that apart from their sign restrictions, the coupling constants $a$ and $b$ are completely free in the above expansion.

\subsubsection{$h(x)$ when $x \rightarrow \pm\infty$}
Another regime where one can find an approximate solution to $h(x)$ is when $x \rightarrow \pm\infty$. It is given by
\begin{align}
&\text{When $x\rightarrow \infty:$} \quad h(x) \sim \frac{1}{\sqrt{1+a^2}}\,e^{a\,x}\,, \nn\\
& \text{When $x\rightarrow -\infty:$} \quad h(x) \sim \frac{1}{\sqrt{1+b^2}}\,e^{b\,x}\,,
\end{align}
where we are taking $a>0$ and $b<0$, but with $a$ and $b$ being otherwise arbitrary constants.

\section{Force Between Two Non-identical Extremal Black Holes\label{sec4}}

After making a proper coordinate choice, the long-range force between two non-identical black holes is just the gradient of the interaction energy given in \eqref{Mintexp}:
\bea
{\cal F}_{12} = \frac{F_{12}}{|\vec{x}_{12}|^2}\,, \quad F_{12} = Q_1\,Q_2 + {\td Q_1}\,{\td Q_2} - M_1\,M_2 - \Sigma_1\,\Sigma_2\,.
\eea
Specialising to extremal black holes, we can use the definitions in \eqref{MSigsols} to rewrite $F_{12}$ in terms of the variables $x_1,\, x_2$,
\bea
F_{12} = -\, e^{-b\,(x_1 + x_2)}\,{\td Q}_1\,{\td Q}_2\, \Big[ h(x_1)\,h(x_2) + h'(x_1)\,h'(x_2) - e^{a\,(x_1+x_2)} - e^{b\,(x_1+x_2)}\Big]\,,
\label{d3int}
\eea
where the parameters 
$a,b$ in the formula above are independent. Note that the force vanishes if $x_1 = x_2$ by virtue of the mass ODE \eqref{diffeq}. Also for $b=-1/a$ and $h(x)$ given by \eqref{hsolBPS}, the force vanishes for \textit{arbitrary} $x_1$ and $x_2$.

\medskip
\noindent\underline{\textbf{In the neighbourhood of $b=-1/a$}}
\medskip

By perturbatively solving for $h(x)$ one can comment on the sign of the force in the neighbourhood of $b=-1/a$. 
Indeed, using the approximate solution \eqref{hsolpert1} in the neighbourhood of $b=-1/a$ one can show that the force $F_{12}$ up to order $\epsilon$ takes the following rather simple form:
\bea
F_{12} &=& -\,\epsilon\,\frac{a^2\,{\td Q_1}\,{\td Q_2}\,(e^{\hat x_1} - e^{\hat x_2})\,(e^{X_1}-1)^2\,(e^{X_2}-1)^2}{(1+a^2)\,(1-a\,e^{\hat x_1})^2\,(1-a\,e^{\hat x_2})^2} \,\Big(Z(X_2) - Z(X_1) \Big) \,,\nn\\
\text{where} && Z(X_i) \equiv \frac{X_i\, e^{X_i}}{(e^{X_i} -1)^2} - \frac{1}{(e^{X_i} -1)}\,, \quad \text{with} \quad X_i \equiv  {\hat x_i} + \log a\,, \quad {\hat x}_i \equiv \frac{(1+a^2)\,x_i}{a}\,. \nn\\
\eea
It can be seen easily that the derivative of the function $Z(X)$ is strictly $\leq 0$ since
\bea
Z'(X) = \frac{1- \frac{X}{2}\,\coth{\frac{X}{2}}}{2\,\sinh{\frac{X}{2}}^2} \leq 0\,, \quad \text{since} \quad X\,\coth X \geq 1\,, \quad \text{for all $X$}\,,
\eea
implying that $Z(X)$ is \textit{strictly monotonically decreasing}. Hence the combination $(e^{\hat x_1} - e^{\hat x_2})\,(Z(X_2) - Z(X_1))$ is strictly $\geq 0$. Therefore we 
can determine  the sign definiteness of $F_{12}$ in the neighbourhood of $b=-1/a$ when $x_1 \neq x_2$:
\bea
\epsilon > 0: \ F_{12} < 0\,, \quad \epsilon = 0: \ F_{12} = 0\,, \quad \epsilon<0:\ F_{12} > 0\,. 
\eea
This behaviour of the mutual force is very reminiscent of the mutual force in the neighbourhood of $a=1$ in EMD black holes \cite{emdpaper}. 

\medskip
\noindent\underline{\textbf{Force for black holes with charge ratio $Q/{\wtd Q} \sim e^{(a-b)x_0}$ with $a,b$ arbitrary}}
\medskip

Let us now calculate the force using a different perturbative solution for $h(x)$ around $x=x_0$ as given in \eqref{hpertx0}, while keeping the coupling constants $a,b$ arbitrary. Using this solution we find that the force, at the leading order, is given by
\bea
F_{12} &=& \fft{(a-b)\,(1-2\,a\,b - \sqrt{1-8\,a\,b})\,(y_1 - y_2)^2\,\epsilon^2}{4\,a} + {\cal O}(\epsilon^3)\,\nn\\
\text{where} && x_0 + \epsilon\,y_{1,2} = x_{1,2}\,,
\eea
and we have suppressed here (and in what follows) positive factors such as $\tilde Q_1 \tilde Q_2$.
Since $a>0$ and $b<0$, the sign of the force is strictly determined by the term $(1-2\,a\,b - \sqrt{1-8\,a\,b})$. We can easily see that
\bea
(1-2\,a\,b)^2 - (1-8\,a\,b) = 4\,a\,b\,(1+a\,b) \, ,
\eea
which implies that
\bea
b > -1/a: \ F_{12} < 0 \,, \quad b = -1/a: \ F_{12} = 0 \,, \quad b < -1/a: \  F_{12} > 0 \,. \label{Ftrend4D}
\eea
This matches with the behaviour of the force around $b = -1/a + \epsilon$ which was obtained using the perturbation presented earlier. Moreover, moving away from $b$ very close to $-1/a$ we find that when two black holes have charge ratios such that $Q/{\td Q} \sim e^{(a-b)\,x_0}$, then the sign of the mutual force behaves the same way.

\medskip
\noindent\underline{\textbf{Force for other limiting cases of charge ratios with arbitrary $a,b$}}
\medskip

Finally we can estimate the mutual force when $x_1 = x_0$ and $x_2 = \pm\infty$, and when $x_1 = \infty, x_2 = -\infty$. In the first case, the force takes the form
\bea
F_{12} (x_0, x_2) = - \left[ \left(\frac{1-\frac{a}{b}}{1+a^2}\right)^{1/2} -1\right]\,e^{a\,(x_0+x_2)}\,,
\eea
and so we have
\bea
b > -1/a: \ F_{12} < 0 \,, \qquad b = -1/a: \ F_{12} = 0 \,, \qquad b < -1/a: \  F_{12} > 0 \,.
\eea
In the  case  $x_1 = \infty, x_2 = -\infty$ the force takes the form (writing $x_1 = k + \lambda,\ x_2 = -k$ and sending $k \rightarrow \infty$):
\begin{align}
F_{12} = -\frac{(1+a\,b)\,e^{(a-b)k+a\lambda}}{(1+a^2)^{1/2}(1+b^2)^{1/2}} \, .
\end{align}
Thus, we see again the same feature for the sign of $F_{12}$ depending on whether $b$ is greater or less than $-1/a$.

\section{Explicit Black Hole Solutions\label{sec5}}

Four-dimensional black hole solutions to the Lagrangian \eqref{2fieldlag} for special choices of $a$ and $b$ 
have appeared
in the literature before. In particular, the Lagrangian obtained in \cite{popetran} from the $S^7$ reduction of $D=11$ supergravity admits the $a=1,\, b=-1$ and $a=\sqrt{3},\, b=-1/\sqrt{3}$ cases of \eqref{2fieldlag}. In \cite{popetran} if one sets $\varphi_1=-2\phi$, \ $\varphi_2 = \varphi_3=0$, $F^1_{(2)} = F^2_{(2)} = {\sqrt2}\, F$, and $F^3_{(2)} = F^4_{(2)} = \sqrt2\, \wtd F$, then one obtains \eqref{2fieldlag} for $a=1,\, b=-1$. On the other hand, setting $\varphi_1 = \varphi_2 = \varphi_3 = -2\phi/\sqrt{3}$,\  $F^1_{(2)} =2F $, and $F^2_{(2)} = F^3_{(2)} = F^{4}_{(2)} = 2 \wtd F/\sqrt{3}$, one finds the $a=\sqrt{3},\, b=-1/\sqrt{3}$ Lagrangian. The black hole solutions for these special choices can also be found from the general asymptotically AdS 4-dimensional solution in \cite{popetran} after taking the ungauged limit. 

Here we present the general static non-extremal black hole solutions to the two-field theory with $b=-1/a$, where $a$ is arbitrary.  These
solutions can also be found, in different conventions, in \cite{gibmae}:
\bea
ds^2 = -\fft{r^2\, h}{\sqrt{V}}\, dt^2 + \fft{\sqrt{V}}{r^2}\, \Big[ h^{-1}\, dr^2 + r^2 \, d\Omega^2\Big]\,,\label{D4met}
\eea
with
\bea
V &=& r^4\, H^{\ft{4}{1+a^2}}\, \wtd H^{\ft{4a^2}{1+a^2}}\,,\qquad h = 1- \fft{\mu}{r}\,,\nn\\
H&=& 1 + \fft{\mu\, s^2}{r}\,,\qquad \wtd H= 1 + \fft{\mu\, \td s^2}{r}\,,
\eea
where $s=\sinh\delta$ and $\td s=\sinh\td\delta$.  The dilaton is given by
\bea
\phi= \fft{a}{1+a^2}\, \log\Big(\fft{\wtd H}{H}\Big)\,,\label{D4dil}
\eea
and the field strengths are given by
\bea
F= \fft{Q}{r^2\,H^2}\, dt\wedge dr\,,\qquad
\wtd F= \fft{\wtd Q}{r^2 \wtd H^2} \, dt\wedge dr\,.\label{D4Max}
\eea
The charges are given in terms of the parameters $\delta$ and $\td\delta$ by
\bea
Q= \fft{\mu\, s \,c}{\sqrt{1+a^2}}\,,\qquad \wtd Q= \fft{a\,\mu\, \td s\, \td c}{\sqrt{1+a^2}}\,,\label{charges}
\eea
where $c=\cosh\delta$ and  $\td c=\cosh\td\delta$.
The mass of the black hole can be seen to be given by
\bea
M= \mu\, \Big[\fft12 + \fft{s^2 + a^2\, \td s^2}{1+a^2}\Big]\,.\label{mass}
\eea

   Comparing the expressions (\ref{charges}) and (\ref{mass}) with the expressions (\ref{totalcharges2}) and (\ref{totalmass}) in
the initial-value formulation, we see that for a single black hole the previous constants $(c_1,d_1,\td c_1,\td d_1)$ are related to
our parameters $\mu$, $\delta$ and $\td\delta$ in the black hole solutions by
\bea
c_1=\ft14\mu\, e^{2\delta}\,,\qquad d_1= \ft14\mu\, e^{-2\delta}\,,\qquad
\td c_1=\ft14\mu\, e^{2\td\delta}\,,\qquad \td d_1= \ft14\mu\, e^{-2\td\delta}\,.
\eea
In the extremal limit, which is given by sending $\mu$ to zero and at the same time sending $\delta$ and $\td\delta$ to infinity,
while keeping $\mu \, s^2$ and $\mu\,\td s^2$ finite and non-zero, the charges become
\bea
Q\longrightarrow \fft{\mu\, s^2}{\sqrt{1+a^2}}\,,\qquad
\wtd Q \longrightarrow \fft{a\,\mu\, \td s^2}{\sqrt{1+a^2}}\,,
\eea
and therefore in this limit the mass becomes
\bea
M_{\rm ext} = \fft{Q + a\, \wtd Q}{\sqrt{1+a^2}}\,.
\eea
This agrees with the expression that we derived in eqn (\ref{extremalmass}) by applying the scaling argument.

\section{Generalization to Higher Dimensions\label{sec6}}

\subsection{Initial data: masses, electric and scalar charges}

The time-symmetric initial-value formulation for the Einstein-Two-Maxwell-Dilaton (E2MD) theory described by the Lagrangian
(\ref{2fieldlag}) was obtained in \cite{kanpop}. Defining the spatial dimension to be $d$, so that the total spacetime dimension is $(d+1)$, it was found that time-symmetric initial
data could be found if the dilaton coupling constants $a$ and $b$ satisfy
\bea
a\, b = - \fft{2(d-2)}{d-1}\,.\label{abrel}
\eea
It is often convenient to parameterise $a$ and $b$ in terms of quantities $N$ and $\wtd N$, such that
\bea
a^2 &=& \fft{4}{N} - \fft{2(d-2)}{d-1}\,,\qquad b^2= \fft{4}{\wtd N} - \fft{2(d-2)}{d-1}\,.\label{abparam}
\eea
The condition (\ref{abrel}) then implies that
\bea
a\, N + b\, \wtd N =0\,,\qquad N +\wtd N = \fft{2(d-1)}{d-2}\,.\label{NtNrels}
\eea

After adjusting the normalisations to suit our conventions in this paper, the ansatz for the time-symmetric initial data is
\bea
\Phi&=& (C\, D)^{\ft{(d-2) N}{4(d-1)}}\, (\wtd C\, \wtd D)^{\ft{(d-2) \wtd N}{4(d-1)}}\,,
\qquad e^\phi = \Big(\fft{C\, D}{\wtd C\, \wtd D}\Big)^{-\ft14 a\, N}\,,
\nn\\
E_\alpha &=& \ft12\sqrt{N}\, 
\Big(\fft{C\, D}{\wtd C\, \wtd D}\Big)^{-\ft{(d-2) \,\wtd N}{2(d-1)}}\,\del_\alpha
\log\Big(\fft{C}{D}\Big)\,,\nn\\
\wtd E_\alpha &=& \ft12\sqrt{\wtd N} \,
\Big(\fft{\wtd C\, \wtd D}{C\, D}\Big)^{-\ft{(d-2)\, N}{2(d-1)}}\,\del_\alpha
\log\Big(\fft{\wtd C}{\wtd D}\Big)\,,\label{initiald}
\eea
where the $d$-metric on the initial time slice is written as
\bea
ds_d^2 = g_{\alpha\beta}\, dx^\alpha\, dx^\beta= \Phi^{\ft4{d-2}}\, dx^\alpha\, dx^\alpha\,.
\eea
The data (\ref{initiald}) will satisfy the initial-value constraints if $C$, $D$, $\wtd C$ and $\wtd D$ are arbitrary 
harmonic functions in the $d$-dimensional flat metric $dx^\alpha\, d x^\alpha$.  Thus, we may take
\bea
C&=& 1 + \sum_{i=1}^{\cal N} \fft{c_i}{|\vec x- \vec x_i|^{d-2}}\,,\qquad
D= 1+\sum_{i=1}^{\cal N} \fft{d_i}{|\vec x- \vec x_i|^{d-2}}\,,\nn\\
\wtd C&=& 1 + \sum_{i=1}^{\cal N} \fft{\td c_i}{|\vec x- \vec x_i|^{d-2}}\,,\qquad
\wtd D=1+ \sum_{i=1}^{\cal N} \fft{\td d_i}{|\vec x- \vec x_i|^{d-2}}\,,
\eea
where $c_i, d_i, \td c_i$ and $\td d_i$ are arbitrary constants.  These
time-symmetric initial data describe a system of ${\cal N}$ black holes, located
at the points $\vec x_i$.

The definition of the ADM mass in an asymptotically-flat $(d+1)$-dimensional spacetime with spatial metric
$ds_d^2= \td g_{\alpha\beta}\, dx^\alpha\, dx^\beta$, where $\td g_{\alpha\beta}=\delta_{\alpha\beta} + h_{\alpha\beta}$
and $h_{\alpha\beta}= {\cal O}(\rho^{-(d-2)}$) with $\rho=\sqrt{x^\alpha\, x^\alpha}$, is conventionally taken to be (see, for example, \cite{eihulesc,jau}),
\bea
M_{\rm ADM}= \fft1{2(d-1)\, \Omega_{d-1}}\,\lim_{\rho\to \infty} \int (\del_\beta h_{\alpha\beta} - 
\del_\alpha h_{\beta\beta})\, n^\alpha\, dS\,,\label{ADMdefd}
\eea
where $n^\alpha = \dfft{x^\alpha}{\rho}$. Here $dS$ is the volume element of the spherical surface at radius $\rho$,
so 
\bea
dS= \Phi^{\ft{2(d-1)}{d-2}} \, \rho^{d-1}\, d\Omega_{d-1}\,,
\eea
with $d\Omega_{d-1}$ being the volume element of the unit $(d-1)$-sphere, whose total volume is $\Omega_{d-1}=\int d\Omega_{d-1}$.  Thus, as in \cite{kanpop}, the total ADM mass $M$ is given in terms of the $\rho^{-(d-2)}$ coefficient in the large-distance expansion of $\Phi$, namely
\bea
\Phi= 1 + \fft{M}{2\rho^{d-2}} + \cdots\,,\label{totalMdefd}
\eea
We therefore find
\bea
M= \fft{d-2}{2(d-1)}\, \sum_i\Big[ (c_i+d_i)\, N + (\td c_i + \td d_i)\, \wtd N\Big]\,.\label{totalMd}
\eea
The total charges $Q$ and $\wtd Q$ will be given by
\bea
Q= \fft1{\Omega_{d-1}}\, \int e^{-2a\phi}\, E_\alpha\, n^\alpha\, dS_{d-1}\,,\qquad
\wtd Q= \fft1{\Omega_{d-1}}\, \int e^{-2b\phi}\, \wtd E_\alpha\, n^\alpha\, dS_{d-1}\,,
\eea
where $dS_{d-1}= \rho^{d-1}\, d\Omega_{d-1}$.  Thus, by evaluating the integrals at large $\rho$ we find
\bea
Q= \ft12 (d-2) \,\sqrt{N}\, \sum_i (c_i-d_i)\,,\qquad
\wtd Q= \ft12  (d-2)\, \sqrt{\wtd N}\, \sum_i (\td c_i - \td d_i)\,.\label{totalchargesd}
\eea

  Finally, we may define the scalar charge $\Sigma$ to be given by
\bea
\Sigma= -\fft1{\Omega_{d-1}}\, \lim_{\rho\to\infty} \int n^\alpha\, \del_\alpha \phi\, dS\,.\label{Sigmaddef}
\eea
Thus, $\Sigma$ is related to the coefficient of the  $\rho^{-(d-2)}$ term in the large-$\rho$ expansion of the
dilaton field $\phi$, 
\bea
\phi=\fft{\Sigma}{(d-2)\, \rho^{d-2}} + \cdots\,.\label{dilatonSig}
\eea
We then find
\bea
\Sigma= -\ft14 a\,N\, (d-2)\, \sum_i(c_i+d_i -\td c_i -\td d_i)\,.\label{totalSigd}
\eea

As in four dimensions, we may calculate the mass of the $i$'th black hole by inverting the radial coordinate 
$\rho\equiv |\vec x - \vec x_i|$ according to 
\bea
\rho'= \fft{\alpha_i^2}{\rho}\,,
\eea
and then reading off $M_i$ from the coefficient of ${\rho'}^{-(d-2)}$ in the large-$\rho'$ expansion of $\Phi$, 
\bea
\Phi = 1+ \fft{M_i}{2{\rho'}^{d-2}} + \cdots\,.
\eea

Again, as in four dimensions, in order for $M_i$ to agree with the expression for the mass of a single black hole as calculated using eqn 
(\ref{totalMdefd}) in the large-$\rho$ region, we must have
\bea
c_i\, d_i = \td c_i\, \td d_i \qquad \hbox{and hence}\qquad \alpha_i= (c_i\, d_i)^{\ft1{2(d-2)}} = (\td c_i\, \td d_i)^{\ft1{2(d-2)}} \, , 
\label{selfd}
\eea
for each value of $i$.  The masses $M_i$ then turn out to be 
\bea
M_i = \fft{d-2}{2(d-1)}\, \Big[ (1+A_i)\, N\, d_i + (1+B_i)\, N\, c_i +
                                (1+\wtd A_i)\, \wtd N\, \td d_i + (1+\wtd B_i)\, \wtd N\, \td c_i\Big]\,,\label{Midefd}
\eea
where
\bea
A_i &=& \sum_{j\ne i} \fft{c_j}{|\vec x_{ij}|^{d-2}}\,,\qquad
B_i=\sum_{j\ne i} \fft{d_j}{|\vec x_{ij}|^{d-2}}\,,\nn\\
\wtd A_i &=& \sum_{j\ne i} \fft{\td c_j}{|\vec x_{ij}|^{d-2}}\,,\qquad
\wtd B_i=\sum_{j\ne i} \fft{\td d_j}{|\vec x_{ij}|^{d-2}}\,.
\eea
The electric charges of the $i$'th black hole, and its scalar charge, can likewise be calculated by inverting through its
location $\vec x_i$. Indeed, this gives the following expressions for the  electric charges
\bea
Q_i &=& \ft12 (d-2)\, \sqrt{N}\, \Big[(1+A_i)\, d_i - (1+B_i)\, c_i\Big]\,,\nn\\
\wtd Q_i &=& \ft12 (d-2)\, \sqrt{\wtd N}\, \Big[(1+\wtd A_i)\, \td d_i - (1+\wtd B_i)\, \td c_i\Big]\,,\label{chargesid}
\eea
and for the scalar charges
\bea
\Sigma_i = -\ft14\, a\, N\, (d-2) \, 
\Big[(1+A_i)\, d_i + (1+B_i)\, c_i - (1+\wtd A_i)\, \td d_i -(1+\wtd B_i)\, \td c_i\Big]\,.
\label{scalarchargeid}
\eea

\subsection{Interactions}

We shall be interested in the interactions between two black holes, and so we set ${\cal N}=2$.  At the leading order, where 
the terms of order $|\vec x_{12}|^{-(d-2)}$ in the right-hand sides of the expressions (\ref{Midefd}), (\ref{chargesid}) and (\ref{scalarchargeid}) are neglected (\ie by neglecting the terms of order $|\vec x_{12}|^{-(d-2)}$ in 
$A_i={\cal O}\big(|\vec x_{12}|^{-(d-2)}\big)$, etc.), we may solve the equations (\ref{Midefd}), (\ref{chargesid}) and 
(\ref{scalarchargeid}) for the eight parameters $(c_1,c_2,d_1,d_2,\td c_1,\td c_2,\td d_1,\td d_2)$ in terms of the eight
quantities $(M_1,M_2,Q_1,Q_2,\wtd Q_1, \wtd Q_2, \Sigma_1,\Sigma_2)$ (see section \eqref{Intenergy4D} for the same calculation in four dimensions):
\bea
c_i &=& \ft12 M_i - \fft{Q_i}{(d-2)\, \sqrt{N}} -
\fft{a\, \Sigma_i}{2 (d-2)}\,,\nn\\
d_i &=& \ft12 M_i + \fft{Q_i}{(d-2)\, \sqrt{N}} -  
\fft{a\, \Sigma_i}{2 (d-2)}\,,\nn\\
\td c_i &=& \ft12 M_i - \fft{\wtd Q_i}{(d-2)\, \sqrt{\wtd N}} - 
              \fft{b\, \Sigma_i}{2 (d-2)}\,,\nn\\
\td d_i &=& \ft12 M_i + \fft{\wtd Q_i}{(d-2)\, \sqrt{\wtd N}} - 
   \fft{b\, \Sigma_i}{2 (d-2)}\,.\nn\\
\label{cdsolsd}
\eea

Substituting the solutions (\ref{cdsolsd}) into the conditions (\ref{selfd}) ensuring the absence of black-hole self energy, and specialising to two identical black holes, we find the condition
\bea
\wtd N\, Q^2 - N\, \wtd Q^2 + a\, (d-1)\, N\, M\, \Sigma +  (N-\wtd N)\, \Sigma^2=0\,.\label{selfdphys}
\eea
As in four dimensions, we may define the interaction energy to be $M_{\rm int}= M- \sum_i M_i$.  We now find that this is given by 
\bea
M_{\rm int} = -\fft{d-2}{(d-1) |\vec x_{12}|^{d-2}}\, \Big[(c_1 \, d_2 +c_2\, d_1)\, N + (\td c_1\,\td d_2 + \td c_2\, \td d_1)\, \wtd N\Big]\,.
\label{Mintres1d}
\eea
Inserting the solutions (\ref{cdsolsd}) into this, we find that for two non-identical black holes
\bea
M_{\rm int} = -\fft{2}{(d-1)(d-2)\,|\vec x_{12}|^{d-2}}\, \Big[\ft12 (d-1)(d-2)\, M_1\, M_2- Q_1\, Q_2 -\wtd Q_1\, \wtd Q_2 
 + \Sigma_1\, \Sigma_2\Big]. \;\label{Mintdunequal}
 \eea
 
 \subsection{Extremal black holes and vanishing interaction energy}
 
 Specialising eqn (\ref{Mintdunequal}) to the case of two identical black holes, we find that the condition
$M_{\rm int}=0$ becomes
\bea
\ft12 (d-1)(d-2)\, M^2 -Q^2 -\wtd Q^2 + \Sigma^2=0\,.\label{extremald}
\eea
This can be understood as the condition for the black hole to be extremal.  

Eliminating $\Sigma$ between the two equations (\ref{selfdphys}) and (\ref{extremald}) implies  
\bea
(d-1)^2\, M^2 - (N\, Q^2 + \wtd N\, \wtd Q^2) = \pm 2\sqrt{N\,\wtd N}\, Q\, \wtd Q\,,
\eea
and hence we obtain the four roots
\bea
(d-1)\, M= \pm\sqrt{N}\, Q \pm \sqrt{\wtd N}\, \wtd Q\,.\label{bpscondsd}
\eea
(The two $\pm$ choices are independent, hence the four roots.)  These are the analogues of the four roots we found in four dimensions in eqns (\ref{BPSconds}). 
Taking, for example, the case where we choose the two plus signs in eqn (\ref{bpscondsd}), the scalar charge is then given by
\bea
\Sigma= \fft{[(d-2)(N-\wtd N) -2(d-1)]\, \sqrt{N}\, Q+
[(d-2)(N-\wtd N) +2(d-1)]\, \sqrt{\wtd N}\, \wtd Q}{2a\,(d-1)\, N}\,. \label{scalchd}
\eea
Note that, just as in four dimensions, the formula for the extremal mass \eqref{bpscondsd} and scalar charge \eqref{scalchd} imply that the interaction energy \eqref{Mintdunequal} for any two non-identical extremal black holes satisfying $a\,b=-\frac{2(d-2)}{d-1}$ should be zero.

\subsection{Mass formula for extremal black holes\label{secheqnd}}

We now derive the higher-dimensional analogue of the mass formula obtained in section 2. 
The scaling argument based on the constant dilaton-shift symmetry is unchanged in higher dimensions. 
The global symmetry of the E2MD Lagrangian (\ref{2fieldlag}) that we 
discussed in section \ref{massform4} in the
context of four dimensions exists also in general dimensions.  Thus, eqns (\ref{globalsym}) - (\ref{massdiff}) are identical
in a general dimension $D=d+1$. 
The only new ingredient is the normalization of the scalar charge 
$\Sigma$ 
relative to $\dfft{\partial M}{\partial \phi_0}$, which depends on the higher-dimensional ADM conventions.
Indeed, in higher dimensions 
we need to determine the precise relation
between $\Sigma$, as defined in eqn (\ref{Sigmaddef}), and $\dfft{\del M}{\del\phi_0}$, where $\phi_0$ is
the asymptotic value of the dilaton field. 
To fix this normalization, we use the standard covariant phase-space derivation of the first law.
We only need the contribution from the boundary at infinity, and in particular only the terms involving the ADM mass and the variation of the asymptotic dilaton.

Viewing the Lagrangian as a $D=d+1$-form $L$, its variation with respect to a field, denoted generically by
$\psi$, takes the form
\bea
\delta L= E(\psi)\cdot\delta\psi + d\Theta(\psi,\delta\psi)\,,
\eea
where $E(\psi)$ is the equation of motion for $\psi$, and $\Theta(\psi,\delta\psi)$ represents the
boundary terms arising from integration by parts.  On shell, $E(\psi)=0$. Taking $\delta\psi$ to be
generated  by a smooth vector field $\xi$ we have 
\bea
\delta L={\cal L}_\xi \, L= (d\,i_\xi + i_\xi\, d)L = d\, i_\xi\, L = d\Theta\,,
\eea
and so we can write 
\bea
\Theta(\psi,{\cal L}_\xi\, \psi)-i_\xi\, L=dQ\,.
\eea
If parameters in the black hole solution are varied we have
\bea
d\delta Q= \delta\Theta - i_\xi\, \delta L = \delta\Theta - i_\xi\, d\Theta =
\delta\Theta -{\cal L}_\xi\,\Theta + d\,i_\xi\, \Theta\,.
\eea
Taking $\xi$ to be a Killing vector gives ${\cal L}_\xi\,\Theta=0$ and $\delta\Theta=0$ (since 
$\delta\Theta =\delta\Theta(\psi,{\cal L}_\xi\,\psi)$ and ${\cal L}_\xi\,\psi=0$).  Hence
$d(\delta Q-i_\xi\,\Theta)=0$ and so $\delta{\cal H}_\infty = \delta{\cal H}_+$, where
\bea
\delta{\cal H}_\infty = \int_\infty (\delta Q-i_\xi\, \Theta)\,,\qquad
\delta{\cal H}_+ = \int_{H_+}  (\delta Q-i_\xi\, \Theta)\,,
\eea
and the integrals are performed over the boundary at infinity and over the outer horizon, respectively.
The equation $\delta{\cal H}_\infty = \delta{\cal H}_+$ yields the first law of black-hole dynamics.  Here, we are concerned just with the terms $\delta M$ and the one coming from varying the asymptotic value
of the dilaton field, and both of these arise in the $\delta {\cal H}_\infty$ integral.  We have 
$\Theta=\Theta^{\rm gravity} + \Theta^{\rm em fields} + \Theta^{\rm dilaton}$, and specifically, we have
\bea
i_\xi\, \Theta^{\rm gravity} &=& -*(\xi\wedge\Xi)\,,\qquad\Xi^\mu= (g^{\mu\alpha}\, g^{\mu\beta} -
g^{\alpha\beta}\, g^{\mu\nu})\,\nabla_\nu\, \delta g_{\alpha\beta}\,,\label{gravterm}\\
i_\xi\, \Theta^{\rm dilaton} &=& 4\delta\phi\, i_\xi * d\phi\,.\label{dilatonterm}
\eea
Taking $\xi$ to be the timelike Killing vector $\dfft{\del}{\del t}$, it follows from the definition 
(\ref{ADMdefd}) for the ADM mass and the definition (\ref{Sigmaddef}) for the scalar charge, that
under a variation of the asymptotic value $\phi_0$ of the dilaton we shall have
\bea
2(d-1)\, \delta M_{\rm ADM}= 4 \Sigma\, \delta\phi_0\,,
\eea
and that therefore the scalar charge can be written as
\bea
\Sigma = \ft12 (d-1)\, \fft{\del M}{\del\phi_0}\,.\label{Sih=gma from M}
\eea
Using eqn (\ref{dhargedphi0}), which continues to be true in arbitrary dimensions, we therefore have
\bea
\Sigma=-\ft12 (d-1)\, \Big[a\, Q\, \fft{\del M}{\del Q} + b\, \wtd Q\, \fft{\del M}{\del \wtd Q}\Big]\,.\label{Sigmaresd}
\eea

We can now proceed in close analogy to the four-dimensional derivation in section \ref{massform4}, 
and impose the extremality condition
(\ref{extremald}).
Using the expression (\ref{Sigmaresd}),  as well as eqn (\ref{massdiff}) to replace the 
$\dfft{\del}{\del\wtd Q}$ derivative by terms involving only the $\dfft{\del}{\del Q}$ derivative, we arrive at an ordinary differential equation for the mass, by defining
\bea
Q= e^{2\lambda\, x}\, \wtd Q\,,\qquad M=\sqrt{Q\,\wtd Q}\, e^{-\mu\,x}\, h(x)\,,\qquad \hbox{with}\quad \mu=\fft{a+b}{2}\,,\quad 
\lambda= \fft{a-b}{2}\,.\label{QMd}
\eea
Thus, we find that $h(x)$ should satisfy the first-order equation
\bea
\ft14(d-1)^2\, {h'}^2 + \ft12 (d-1)(d-2)\, h^2 = e^{2a\, x} + e^{2b\, x}\,.\label{heqnd}
\eea
The mass $M$ and the electric charge $Q$ are then given in terms of $\wtd Q$, and $h(x)$ and $x$, by eqns (\ref{QMd}), and so
\bea
M= \wtd Q\, e^{ -b\, x}\,h(x)\,,\qquad Q= \wtd Q\, e^{(a-b)\, x}\,.\label{MassQd}
\eea
The scalar charge is given by
\bea
\Sigma= -\ft12 (d-1)\, \wtd Q\, e^{-b\, x}\, h'(x)\,.\label{Sigmad}
\eea

It should be emphasised that our results in subsection \ref{secheqnd} apply to extremal black holes for arbitrary values of the
two independent dilaton coupling constants $a$ and $b$.  In the next subsection, we consider explicit 
black hole solutions, which can be constructed specifically in the cases where $a$ and $b$ are related 
as in eqn (\ref{abrel}).

In the special case where $a$ and $b$ are related by eqn (\ref{abrel}), we can use the results we obtained previously from the initial-value formulation in order to find the explicit solution to the equation (\ref{heqnd}).  Thus, substituting $M$ and $Q$ given in eqns (\ref{MassQd}) into eqn (\ref{bpscondsd}), we see that $h(x)$ will be given by
\bea
h(x) = \fft1{d-1}\, \Big(\sqrt{N}\, e^{a\, x} + \sqrt{\wtd N}\, e^{b\, x}\Big)\,.\label{hsold0}
\eea
It can easily be verified that this expression for $h(x)$ indeed satisfies eqn (\ref{heqnd}), provided that the condition (\ref{abrel}) holds.

\subsection{Explicit black hole solutions\label{hdsolns}}

As a consistency check, in this subsection we may test the mass formula (\ref{hsold0}) we just derived, by comparing it with results for some known black hole solutions.
Explicit non-extremal black hole solutions to the $D=d+1$ dimensional E2MD theory (\ref{2fieldlag}) were constructed in \cite{gibmae}, specifically for the case where the dilaton coupling constants obey the relation (\ref{abrel}), and therefore also eqns (\ref{NtNrels}) hold.  A more convenient presentation of the solutions appeared in \cite{lu2field}, and also in \cite{kanpop}.  

Adapted to our conventions, the non-extremal static black hole solutions are given by
\bea
ds^2 &=& -(H^N\, \wtd H^{\wtd N})^{-\ft{d-2}{d-1}}\, f\, dt^2 + (H^N\, \wtd H^{\wtd N})^{\ft{1}{d-1}}\,
\big( f^{-1}\, dr^2 + r^2\, d\Omega_{d-1}^2\big)\,,\nn\\
A&=& \fft{\sqrt{N}\, c}{2s}\, H^{-1}\, dt\,,\qquad
\wtd A= \fft{\sqrt{\wtd N}\, \td c}{2\td s}\, \wtd H^{-1}\, dt\,,\nn\\
\phi&=& -\ft14 a\, N\, \log H - \ft14 b\, \wtd N\, \log \wtd H\,,\qquad
f= 1 - \fft{\mu}{r^{d-2}}\,,\nn\\
H &=& 1 + \fft{\mu\, s^2}{r^{d-2}}\,,\qquad \wtd H = 1 + \fft{\mu\, \td s^2}{r^{d-2}}\,,\label{bhd1}
\eea
where we have defined $c=\cosh\delta$,\  $s=\sinh\delta$,\ $\td c=\cosh\td\delta$ and $\td s =\sinh\td\delta$.  The constant
$\mu$ is a parameter related to the mass, while the parameters $\delta$ and $\td\delta$ are related to the electric
charges $Q$ and $\wtd Q$.  Concretely, we have
\bea
M &=& \fft{\mu}{2}\, \Big[1 + \fft{d-2}{d-1}\, (N\, s^2 + \wtd N\, \td s^2)\Big]\,,\nn\\
Q &=& \ft12 (d-2)\, \mu\,\sqrt{N}\, s\, c\,,\qquad \wtd Q= \ft12(d-2)\, \mu\, \sqrt{\wtd N}\, \td s\, \td c\,.
\label{nonextMQtQ}
\eea

Making the coordinate transformation from $r$ to $\rho$, where
\bea
r^{d-2}= \rho^{d-2}\, \Big(1 + \fft{\mu}{4 \rho^{d-2}}\Big)^2\,,\label{rtorho}
\eea
casts the metric into the isotropic form
\bea
ds^2 = -\Big(1-\fft{\mu^2}{16\rho^{2(d-2)}}\Big)^2 \,\Big[(C\,D)^N\, (\wtd C\,\wtd D)^{\wtd N}\Big]^{-\ft{d-2}{d-1}}\, dt^2 +\Phi^{\ft4{d-2}}\, (d\rho^2 + \rho^2\, d\Omega_{d-1}^2)\,,
\eea
where
\bea
\Phi =(C\, D)^{\ft{(d-2)\, N}{4(d-1)}}\, (\wtd C\, \wtd D)^{\ft{(d-2)\, \wtd N}{4(d-1)}}\,,\label{PhiCD}
\eea
and
\bea
C &=& 1 + \fft{\mu\, e^{2\delta}}{4\rho^{d-2}}\,,\qquad
D=1 + \fft{\mu\, e^{-2\delta}}{4\rho^{d-2}}\,,\nn\\
\wtd C &=& 1 + \fft{\mu\, e^{2\td\delta}}{4\rho^{d-2}}\,,\qquad
\wtd D=1 + \fft{\mu\, e^{-2\td\delta}}{4\rho^{d-2}}\,.\label{CDCCDD}
\eea
It can be seen that the expression (\ref{PhiCD}) is precisely of the form given in the initial data (\ref{initiald}), with
$C$, $D$, $\wtd C$ and $\wtd D$ in eqns (\ref{CDCCDD}) being harmonic functions providing initial data for a single black hole.

The extremal limit corresponds to sending the blackening parameter $\mu$ to zero, while simultaneously 
sending the charge parameters $\delta$ and $\td\delta$ to infinity, such that $q$ and $\td q$ 
\bea
q=\mu\, s^2\qquad \hbox{and} \qquad \td q= \mu\, \td s^2
\eea
are held fixed.  It therefore follows that in the extremal limit, the mass and the electric charges given in eqns (\ref{nonextMQtQ}) 
become
\bea
M &=& \fft{d-2}{2(d-1)}\, \big( N\, q + \wtd N\, \td q\big)\,,\nn\\
Q &=& \ft12 (d-2)\, \sqrt{N}\, q\,,\qquad 
\wtd Q= \ft12 (d-2)\, \sqrt{\wtd N}\, \td q\,.
\eea
From the previous definitions it can then be seen that the mass function $h(x)$ is given by
\bea
h(x) = \fft1{d-1}\, \big( \sqrt{N}\, e^{ax} + \sqrt{\wtd N}\, e^{bx}\big)\,,\label{hsold}
\eea
which indeed agrees with the previous result (\ref{hsold0}), which we obtained from the initial-value formulation.

\subsection{Interaction energy and force between extremal black holes}

The discussion of the force between non-identical extremal black holes that we gave earlier in four dimensions can be generalised to the $D=d+1$ dimensional cases for arbitrary $d$.  In fact this can be done without any additional work, by noting that the differential equation (\ref{heqnd}) 
that we obtained for the mass of extremal black holes in $D=d+1$ dimensions can be mapped into the equation (\ref{diffeq})
we derived in four dimensions.  We do this by starting from eqn (\ref{heqnd}), and then defining
\bea
h(x)= \alpha\, \hat h(x)\,,\qquad x=\beta\, \hat x\,,\qquad a =\beta^{-1}\, \hat a\,,\qquad b= \beta^{-1}\,\hat  b\,.\label{rescale}
\eea
If the constants $\alpha$ and $\beta$ are chosen such that
\bea 
\alpha^2= \fft{2}{(d-1)(d-2)}\,,\qquad \beta^2= \fft{d-1}{2(d-2)}\,,\label{albe}
\eea
then it can be seen that eqn (\ref{heqnd}) becomes
\bea
\hat h'(\hat x)^2 + \hat h(\hat x)^2 = e^{2\hat a\,\hat x} + e^{2\hat b\, \hat x}\,,
\eea
precisely of the form of the $d=3$ equation (\ref{diffeq}).  Thus, the solution for the mass function $h(x)$ for general $d$ can be obtained from the corresponding solution for $d=3$, by applying the rescalings to the hatted variables.  Note that if one specialises to the case where the dilaton couplings $a$ and $b$ obey the condition (\ref{abrel}), then the rescaled couplings $\hat a$ and $\hat b$ obey the condition $\hat a\, \hat b=-1$, which is precisely the $d=3$ relation seen in eqn (\ref{barel}).

The formulae for the mass $M$ and electric charge $Q$ in eqns (\ref{MassQd}), and the scalar charge $\Sigma$ in eqn (\ref{Sigmad}), become
\bea
M= \alpha\, \wtd Q\, e^{-\hat b\,\hat x}\, \hat h(\hat x)\,,\qquad Q= \wtd Q\, e^{(\hat a-\hat b)\,\hat x}\,,\qquad
\Sigma= -\wtd Q\, e^{-\hat b\,\hat x}\, \fft{d\hat h(\hat x)}{d\hat x}
\eea
in terms of the hatted variables.   Thus from eqn (\ref{Mintdunequal}) the interaction energy between two unequal extremal black
holes in $D=d+1$ dimensions becomes
\bea
M_{\rm int} = -\fft{2\wtd Q_1\, \wtd Q_2\, e^{-\hat b(\hat x_1+\hat x_2)}}{(d-1)(d-2)\,|\vec x_{12}|^{d-2}}\, 
\Big[\hat h_1\, \hat h_2 + \hat h_1'\, \hat h_2' - e^{\hat a\,(\hat x_1+\hat x_2)} - e^{\hat b\, (\hat x_1+\hat x_2)} \Big]\,,\label{Mintdunequal2}
\eea
where $\hat h_i=\hat h(\hat x_i)$ and $\hat h_i' = \dfft{d\hat h(\hat x)}{d\hat x}\Big|_{\hat x= \hat x_i}$.  Since this is proportional 
to the four-dimensional expression (\ref{d3int}) for the interaction energy, it follows that the previous perturbative proofs we gave for the interaction energies in various regimes in the four-dimensional case extend immediately to analogous results in all higher dimensions.   Thus, the force $F_{12}$ will be such that depending on the relative magnitudes of the dilaton couplings $a$ and $b$,
\bea
b + \fft{2(d-2)}{(d-1)\, a} >0:&& \qquad F_{12} <0\,,\nn\\
b + \fft{2(d-2)}{(d-1)\, a} =0:&& \qquad F_{12} =0\,,\nn\\
b + \fft{2(d-2)}{(d-1)\, a} <0:&& \qquad F_{12} >0\,,
\eea
showing a clear demarcation between attractive and repulsive behavior, based purely on the range of dilaton couplings.

\section{Conclusions}

One of our motivations in this paper was to explore the behavior of the 
long-range interaction (\ref{introforce}) between non-identical black holes in the theory (\ref{introlag}), 
and in particular 
ask whether the repulsive or attractive nature of the force is correlated with a particular regime of parameter space of the theory.

Indeed, a central result of our analysis is that the special parameter choice
\begin{equation}
b=-{1\over a}
\end{equation}
acts as a zero-force, BPS-like condition, separating an attractive region from a repulsive region in the space of dilaton couplings.
On the special locus $b=-\dfft1a$ the time-symmetric initial-value construction is available, the extremal mass formula is explicit, and the force between any two extremal black holes vanishes, even when the black holes carry different charges. In the regimes analyzed in this work, and assuming $a>0$, we find
\begin{equation}
\label{forceconclusions}
b>-{1\over a} \;\; \Longrightarrow \;\; F_{12}<0\, ,
\quad
b=-{1\over a} \;\; \Longrightarrow \;\; F_{12}=0 \, ,
\quad
b<-{1\over a} \;\; \Longrightarrow \;\; F_{12}>0 \, .
\end{equation}

We also derived a mass equation for extremal black holes with arbitrary values of the two dilaton couplings. In four dimensions this equation takes the form
\begin{equation}
h'(x)^2+h(x)^2=e^{2 a x}+e^{2 b x},
\end{equation}
where the function $h(x)$ encodes the extremal mass, electric charges and scalar charge. Although this equation does not appear to admit a closed-form solution for generic $a$ and $b$, it provides a useful way of studying the force between non-identical extremal black holes without having the full solution. Perturbative solutions near $b=-1/a$, expansions around the point of vanishing scalar charge, and asymptotic solutions all lead to the same qualitative conclusion: the sign of the force changes as one crosses the special force-cancellation locus.

We also saw that these results can be generalized to higher dimensions.
In $D=d+1$ spacetime dimensions, the special locus on which the initial-value construction exists is
\begin{equation}
ab=-{2(d-2)\over d-1}.
\end{equation}
Equivalently, for $a>0$, the relation giving the force-cancellation condition $F_{12}=0$ is 
\begin{equation}
b=-{2(d-2)\over (d-1)a}.
\end{equation}
The sign of the force is then 
\begin{equation}
b+{2(d-2)\over (d-1)a}>0
\;\;\Longrightarrow \;\; F_{12}<0 \,,
\qquad
b+{2(d-2)\over (d-1)a}<0
\;\;\Longrightarrow \;\; F_{12}>0 \,.
\end{equation}
Thus, we see again a BPS-like relation between the couplings, which separates attractive and repulsive regions of parameter space.

Intriguingly, the Toda black holes studied in Appendix B contain branches for which the force condition (\ref{forceconclusions}) is not obeyed. However, we were able to show that such branches are ruled out because they contain naked singularities outside of the horizon (see Appendix B.1 for a detailed discussion). It would be interesting to further explore the (quite unexpected) connection
between regularity of a given black hole solution and
the nature of long-range interactions between two such solutions, to determine whether it is robust.

Another extension of our analysis would be to include higher-derivative corrections to the theory. At the two-derivative level, the vanishing self-force of an extremal black hole is a central ingredient in the mass equation. Once higher-derivative interactions are included, however, the corrected extremality relation and the moduli dependence of the mass can produce a nonzero self-force, even at zero temperature (see e.g. \cite{Cremonini:2021upd}). The resulting mass equation should therefore acquire a source determined by the  Wilson coefficients controlling the higher-derivative terms. It would be particularly interesting to determine whether the two-derivative force-cancellation locus is shifted to a Wilson-coefficient-dependent hypersurface, or whether it disappears altogether, and whether regularity, unitarity or positivity constraints restrict the sign of the corrected force.

It would also be interesting to determine if the sign of the binding energy tracks the sign of the force (as it did in our earlier work on EMD theory \cite{emdpaper}), and whether the attractive/repulsive transition is equivalent to a change in convexity of $M(Q,\tilde{Q})$.

Finally, our work proves the sign rule (\ref{forceconclusions}) in several perturbative and asymptotic regimes. A valuable next step would be to solve the simple first-order mass equation numerically over the full parameter space, scanning for potential counterexamples. This approach could then motivate a global analytic proof.
We leave these questions to future work.

\section*{Acknowledgments}

The work of S.C. was supported in part by the National Science Foundation under Grant No. PHY-2210271. 
The work of M.C. is supported by the DOE (HEP) Award DE-SC0013528, the Slovenian Research Agency (ARRS No. P1-0306) and the Fay R. and Eugene L. Langberg Endowed Chair funds.
The work of C.N.P. is supported in part by the DOE Grant No. DE-SC0010813.
 The authors would like to thank Cook's Branch 2026, where part of this work was carried on, for hospitality and support. C.N.P. is grateful to the Albert Einstein Institute, Potsdam, for hospitality and support.

\appendix

\section{\label{D4con} Extremality condition for black holes for general $a$ and $b$}

\subsection{$D=4$ dimensions}

The ansatz for extremal black hole solutions to the four-dimensional theory described by the Lagrangian (\ref{2fieldlag}) can be written as
\bea
ds_4^2 &=& - H^{-\ft{2}{1+a^2}}\, \wtd H^{-\ft{2}{1+b^2}}\, dt^2 +
  H^{\ft{2}{1+a^2}}\, \wtd H^{\ft{2}{1+b^2}}\, (dr^2 + r^2\, d\Omega_2^2)\,,\nn\\
e^\phi &=& H^{-\ft{a}{1+a^2}}\, \wtd H^{-\ft{b}{1+b^2}}\,,\nn\\
F&=& \fft{Q}{r^2\, H^2}\, \wtd H^{-\ft{2(1+ab)}{1+b^2}}\, dt\wedge dr\,,\qquad
\wtd F= \fft{\wtd Q}{r^2\, \wtd H^2}\, H^{-\ft{2(1+ab)}{1+a^2}}\, dt\wedge dr\,.\label{D4extremalans}
\eea
It can be seen that this ansatz reduces to that in eqns (\ref{D4met}), (\ref{D4dil}) and (\ref{D4Max}) in the extremal case and with the specialisation $b=-\dfft1{a}$.  It can also be seen that, after the appropriate rescalings to match our current conventions, the ansatz in
eqns (\ref{D4extremalans}) reduces to the one given in eqns  (2.1) of \cite{emdpaper}, after  letting $b=-a$ and taking $\mu=0$ in that paper in order to specialise to the extremal case.

For general values of $a$ and $b$, we may substitute the ansatz (\ref{D4extremalans}) into the equations of motion following from eqn (\ref{2fieldlag}).  The Hamiltonian constraint, which follows from eliminating the $H''$ and $\wtd H''$ terms from the Einstein equations,
becomes
\bea
&&\fft1{1+a^2}\, \Big(\fft{H'}{H}\Big)^2+ \fft1{1+b^2}\, \Big(\fft{\wtd H'}{\wtd H}\Big)^2 +
  \fft{2(1+a b)}{(1+a^2)(1+b^2)}\, \fft{H'\,\wtd H'}{H\, \wtd H} \nn\\
&&-
  \fft{Q^2}{r^4\, H^2}\, \wtd H^{-\ft{2(1+a b)}{1+b^2}} - \fft{\wtd Q^2}{r^4\, \wtd H^2}\, H^{-\ft{2(1+a b)}{1+a^2}}=0\,.
  \label{D4hamcon}
  \eea
Substituting large-$r$ expansions of the functions $H$ and $\wtd H$ of the form
\bea
H(r)= 1 + \fft{h_1}{r} + \fft{h_2}{r^2}+\cdots\,,\qquad
\wtd H(r) = 1 +\fft{\td h_1}{r} + \fft{\td h_2}{r^2}+\cdots\,,\label{HHHexps}
\eea
with $h_i$ and $\td h_i$ constants, 
the Hamiltonian constraint (\ref{D4hamcon}) at leading order in $\dfft1{r}$ gives
\bea
\fft{h_1^2}{1+a^2} + \fft{\td h_1^2}{1+b^2} + \fft{2 (1+a b)\, h_1\, \td h_1}{(1+a^2)(1+b^2)} -Q^2-\wtd Q^2=0\,.
\label{D4hamcon2}
\eea
The mass $M$ can be read off from the large-$r$ expansion of the metric in eqn (\ref{D4extremalans}), and the scalar charge $\Sigma$ can be read off from the coefficient of $\dfft1{r}$ in the large-$r$ expansion for $\phi(r)$:
\bea
M= \fft{h_1}{1+a^2} + \fft{\td h_1}{1+b^2}\,,\qquad
\Sigma= -\fft{a\, h_1}{1+a^2} - \fft{b\, \td h_1}{1+b^2}\,.
\eea
Solving for $h_1$ and $\td h_1$ gives
\bea
h_1 = -\fft{(1+a^2)\, (b\, M + \Sigma)}{a-b}\,,\qquad
\td h_1 = \fft{(1+b^2)\, (a\, M+\Sigma)}{a-b}\,.\label{hsol}
\eea
Substituting these back into the Hamiltonian constraint (\ref{D4hamcon2}) yields
\bea
M^2 -Q^2 -\wtd Q^2 + \Sigma^2 =0\,.
\eea
Thus, we have reproduced the extremality condition obtained in eqn (\ref{extremalcon}).  Note, however, that we have derived it here for arbitrary values of the dilaton couplings $a$ and $b$, whereas in the earlier discussion leading up to eqn (\ref{extremalcon}) we had assumed the restriction that $b=-\dfft1{a}$. 

For completeness, we record that the full set of equations that result from substituting the ansatz (\ref{D4extremalans}) into the 
Einstein, Maxwell and dilaton equations following from the Lagrangian (\ref{2fieldlag}) reduce to
\bea
H'' &=& \fft{{H'}^2}{H} -\fft{2 H'}{r} - \fft{(1+a^2)\, Q^2}{r^4\, H} \,\wtd H^{-\ft{2(1+a b)}{1+b^2}}\,,\nn\\
\wtd H'' &=& \fft{\wtd {H}{'}^2}{\wtd H} -\fft{2 \wtd H'}{r} - 
           \fft{(1+b^2)\, \wtd Q^2}{r^4\, \wtd H} \,H^{-\ft{2(1+a b)}{1+a^2}}\,,\label{HppHHpp}
\eea
together with the Hamiltonian constraint (\ref{D4hamcon}).  It is also worth noting that if the expansions (\ref{HHHexps}) are
substituted into eqns (\ref{HppHHpp}), then the leading-order equations at large $r$ give
\bea
h_1^2 - 2 h_2 -(1+a^2)\, Q^2=0\,,\qquad  \td h_1^2 -2 \td h_2 - (1+b^2)\, \wtd Q^2=0\,.
\eea
These are the generalisations of the expressions in eqns (2.7) of \cite{emdpaper} to our present case, where $a$ and $b$ can be arbitrary.

\subsection{General dimensions}

The derivation in this appendix generalises straightforwardly to arbitrary dimensions $D=d+1$.  The ansatz for extremal black holes for unrestricted values of the dilaton couplings $a$ and $b$ takes the form
\bea
ds^2 &=& -\Big( H^{N}\, \wtd H^{\wtd N}\Big)^{-\ft{d-2}{d-1}}\, dt^2 + 
\Big(H^{N}\, \wtd H^{\wtd N}\Big)^{\ft1{d-1}}\,
(dr^2 + r^2\, d\Omega_{d-1}^2)\,,\nn\\
\phi &=& -\ft14 a\, N\, \log H -\ft14 b\, \wtd N\, \log\wtd H\,,\nn\\
F&=& \fft{Q}{r^{d-1}\, H^2}\, \wtd H^{-\ft12 c\,\wtd N} \, dt\wedge dr\,,\nn\\
\wtd F &=& \fft{\wtd Q}{r^{d-1}\, \wtd H^2}\, H^{-\ft12 c\, N} \, dt\wedge dr\,,
\eea
where we have defined 
\bea
c\equiv a b+ \fft{2(d-2)}{d-1}\,,\label{cdef}
\eea
and where we parameterise $a$ and $b$ as
\bea 
a^2= \fft{4}{N} - \fft{2(d-2)}{d-1}\,,\qquad b^2= \fft4{\wtd N} - \fft{2(d-2)}{d-1}\,.
\eea
We note that the restrictions (\ref{abrel}) and (\ref{NtNrels}) are no longer imposed as they were when we were considering the initial value formulation (note that if the condition (\ref{abrel}) {\it is} imposed, then the constant $c$ defined in eqn (\ref{cdef})
vanishes).  

It can now be verified that the Einstein, dilaton and Maxwell equations are all satisfied provided that the functions $H(r)$ and
$\wtd H(r)$ satisfy the second-order equations
\bea
H''&=& \fft{{H'}^2}{H}-\fft{(d-1)\, H'}{r} -
   \fft{4 Q^2}{N\, r^{2d-2}\, H}\, \wtd H^{-\ft12 c\,\wtd N}\,,\nn\\
\wtd H'' &=&  \fft{{\wtd H'{}}^2}{\wtd H} -\fft{(d-1)\, \wtd H'}{r} -
   \fft{4 \wtd Q^2}{\wtd N\, r^{2d-2}\,  \wtd H}\,H^{-\ft12 c\, N}\,,
\eea
together with the first-order Hamiltonian constraint
\bea
&&N\, \Big(\fft{H'}{H}\Big)^2 + \wtd N\, \Big(\fft{\wtd H'}{\wtd H}\Big)^2 +
  \ft12 c\, N\,\wtd N\, \fft{H'\, \wtd H'}{H\,\wtd H} \nn\\
  &&-\fft{4 Q^2}{r^{2d-2}\, H^2}\, {\wtd H}^{-\ft12 c\, \wtd N}
- \fft{4 \wtd Q^2}{r^{2d-2}\, \wtd H^2}\, {H}^{-\ft12 c\, N} =0\,.\label{hamcond}
\eea

At large distance, the functions $H(r)$ and $\wtd H(r)$ have expansions of the form
\bea
H= 1 + \fft{h_1}{r^{d-2}} + \fft{h_2}{r^{d-1}} +\cdots\,,\qquad
\wtd H= 1 + \fft{\td h_1}{r^{d-2}} + \fft{\td h_2}{r^{d-1}} +\cdots\,.
\eea
Substituting these into the constraint equation (\ref{hamcond}) yields, at the leading order $\dfft1{r^{2d-2}}$, the relation
\bea
(d-2)^2\, \big[ N\, h_1^2 + \wtd N\, \td h_1^2 + \ft12 c\, N\, \wtd N\, h_1\, \td h_1\big] -4Q^2 -4 \wtd Q^2=0\,.
\label{cond2}
\eea
The mass can be read off in the usual way from the large-$r$ expansion  of 
$\Phi=\big(H^N\, \wtd H^{\wtd N}\big)^{\ft{d-2}{4(d-1)}}$, 
as in eqn (\ref{totalMdefd}), and the scalar charge $\Sigma$ can be read off from the
$\dfft1{r^{d-2}}$ term in the expansion of the dilaton, as in eqn (\ref{dilatonSig}).  Thus, we have
\bea
M= \fft{d-2}{2(d-1)}\, \big( N\, h_1 + \wtd N\, \td h_1\big)\,,\qquad
\Sigma= -\ft14 (d-2)\, \big( a\, N\, h_1 + b\, \wtd N\, \td h_1\big)\,,
\eea
from which we find
\bea
h_1 =-\fft{2b\, (d-1)\, M + 4\Sigma}{(a-b)\, (d-2)\, N}\,,\qquad
\td h_1 = \fft{2a\, (d-1)\, M + 4\Sigma}{(a-b)\, (d-2)\, \wtd N}\,.
\eea
Finally, substituting these into the constraint equation (\ref{cond2}) we obtain, after some straightforward 
algebraic manipulations, 
\bea
\ft12 (d-1)(d-2)\, M^2 -Q^2 - \wtd Q^2 + \Sigma^2=0\,.\label{extremald2}
\eea
Thus we have reproduced the same extremality condition that we obtained previously in eqn (\ref{extremald}) under the restriction that $a$ and $b$ were related by eqn (\ref{abrel}) (in order that we could make use of the initial-value formulation).  Here, by
contrast, we have obtained the result (\ref{extremald2}) in the more general situation where $a$ and $b$ are arbitrary and
independent.

\section{Toda Black Hole Solutions\label{Todasec}}

   In addition to the class of explicit four-dimensional
black hole solutions with $b=-\dfft1{a}$ that we described in section \ref{sec5}, and their higher dimensional generalisations in section {\ref{hdsolns}}, there are
also certain exceptional cases where explicit solutions can be obtained.  A well known and long standing example is the case where, in four dimensions,  
$a=-b=\sqrt3$, for which the equations for static black holes can be reduced to
a Toda system for the Lie algebra $A_2$ (see, for example, \cite{gibmae} and
references therein).  The $A_2$ Toda solution generalises also to higher 
dimensions.  

In a recent paper \cite{luwuwuzh}, it was shown that the equations for static black holes in the E2MD theory, which can always be cast into the form of a Toda-like system, become exactly solvable also if the dilaton couplings $a$ and $b$ take certain other specific values.  Namely, it is possible to choose the dilaton coupling constants so that the Toda-like system becomes precisely the integrable Toda system of the   $B_2$ or the $G_2$ Lie algebras.  The corresponding non-extremal static black hole solutions can then be obtained explicitly, in arbitrary dimensions \cite{luwuwuzh}.  Here, we shall be interested in the extremal limits.

Since the complete discussion of these Toda black hole solutions is quite involved, here we shall refer to \cite{luwuwuzh} for all the details, and simply quote the results that are necessary for our purposes.   Our goal here is to obtain the solutions to the mass equation (\ref{diffeq}) for extremal black holes corresponding to each of the Toda cases. It suffices for us to discuss just the four-dimensional black holes, since as we already showed, the corresponding mass equation (\ref{heqnd}) in higher dimensions can be mapped into the four-dimensional equation by means of the transformations (\ref{rescale}) and (\ref{albe}).

Looking at our equations (\ref{MSigsols}), it can be seen that we just need to know the ratios $\dfft{M}{\wtd Q}$ and $\dfft{Q}{\wtd Q}$ for a given extremal solution in order to be able to determine the corresponding solution $h(x)$ to the mass equation (\ref{diffeq}).  The normalisation conventions for the electric charges $Q_1$ and $Q_2$  and the mass $M$ that are employed in \cite{luwuwuzh} are significantly different in general dimensions, but in four dimensions the 
normalisations for mass are the same, and for the charges our $Q$ and $\wtd Q$ are simply related to their $Q_1$ and $Q_2$ by $Q= 2 Q_1$ and $\wtd Q=2 Q_2$. 

\bigskip
\noindent\underline{{\bf $A_2$ Toda black hole}}:
\medskip

As already mentioned, this example is well known and of long standing.  The dilaton coupling constants are given by
\bea
a=\sqrt3\,,\qquad b=-\sqrt3 \label{A2ab}\,,
\eea
and in the
normalisation conventions we are using, the mass and the charges for
the extremal $A_2$ black hole are related by
\bea
M = \ft12\, \Big( Q^{\ft23} + \wtd Q^{\ft23}\Big)^{\ft32}\,.
\label{A2mass}
\eea
Using eqns (\ref{MSigsols}), this translates into the statement that
\bea
h(x) = \sqrt2\, \Big(\cosh\fft{2x}{\sqrt3}\Big)^{\ft32}\,.
\eea
It can be verified that this indeed solves our mass equation (\ref{diffeq}),
provided that $a$ and $b$ are given by eqns (\ref{A2ab}).

  From eqn (\ref{d3int}), the mutual force between two non-identical extremal $A_2$ Toda black holes is given by
\bea
\frac{F_{12}}{{\wtd Q}_1 {\wtd Q}_2} = 1 + q_1^3\,q_2^3 - \sqrt{1+q_1^2}\,\sqrt{1+q_2^2}\,\Big[ 1 - \ft12\,(q_1^2+q_2^2) + q_1^2\,q_2^2 \Big]\,,
\eea
where we have defined $q_1= e^{2x_1/\sqrt3}$ and $q_2=e^{2x_2/\sqrt3}$.  If we now define 
$q_1= \tan t_1$ and $q_2=\tan t_2$, where the $t$ parameters range from 0 to $\ft12\pi$, 
then the force can be written as
\bea
\frac{F_{12}}{\wtd Q_1 \wtd Q_2} = \fft{\sin^2\ft12(t_1-t_2)}{4\cos^3 t_1\, \cos^3\, t_2}\, 
S_{12}\,,
\eea
where
\bea
S_{12} &=& 4 \cos(t_1-t_2) - \cos(2t_1-2t_2) - 3\cos(2t_1 + 2 t_2)\,,\\
&=& 6 \sin^2(t_1+t_2) - 8\sin^4\ft12(t_1-t_2)\,.
\eea
This can be seen to be non-negative when plotted.  In fact it can be factorised, and so for $t_1$ and $t_2$ in their allowed ranges
one just needs to prove
\bea
\sqrt3\, \sin(t_1+t_2) - 2 \sin^2\ft12(t_1-t_2) \ge 0\,,
\eea
or, equivalently, that
\bea
\sqrt3\, \sin u + \cos v-1\ge0\,,\label{inequality1}
\eea
where we define $u=t_1+t_2$ and $v=t_1-t_2$.  Note that $0\le u\le \pi$ and $-\ft12\pi\le v\le\ft12\pi$, so $\sin u\ge0$ and $\cos v\ge 0$.  
Also, since $\cos v-\cos u=2\sin t_1\, \sin t_2$ and $\cos v+ \cos u= 2\cos t_1\, \cos t_2$, and both
right-hand sides are non-negative for all allowed $t_1$ and $t_2$, it follows that
$\cos v\ge |\cos u|$.  Consequently 
\bea
\sqrt3\, \sin u+\cos v\ge \sqrt3\, \sin u + |\cos u|\,.\label{bound1}
\eea
For the range $0\le u\le \ft12\pi$ we can write
\bea
 \sqrt3\, \sin u + |\cos u| = \sqrt3\, \sin u + \cos u = 2 \sin\big(u+\fft{\pi}{6}\big)\,,
\eea
while for $\ft12\pi\le u\le\pi$ we can write
\bea
 \sqrt3\, \sin u + |\cos u| = \sqrt3\, \sin u - \cos u = 2 \sin\big(u-\fft{\pi}{6}\big)\,,
\eea
Thus it is evident that $\sqrt3\, \sin u + |\cos u|\ge1$ for the entire range $0\le u\le \pi$, and thus from eqn (\ref{bound1}) it can be seen that the inequality (\ref{inequality1}) is indeed satisfied.

This establishes that the force between two unequal extremal $A_2$ black holes is always non-negative, in line with the general pattern we have exhibited in this paper since for these black holes $b < -1/a$. 

The mass function $h(x)$ for $A_2$ black hole can be straightforwardly extended to higher dimension by using \eqref{rescale}. Since the expression for mutual force in higher dimension \eqref{Mintdunequal2} is identical to the one in four dimension, the non-negativity of the mutual force will continue to hold for $A_2$ black hole sin higher dimensions.

\bigskip
\noindent\underline{\bf {$B_2$ Toda black hole}}:
\medskip

From the results in eqns (61) of \cite{luwuwuzh}, in four dimensions the mass and electric charges for the extremal  $B_2$ Toda black hole will be such that, in our normalisations, 
\bea
\fft{M}{\wtd Q} = 
\fft{9- 3\alpha- \alpha^2}{\sqrt{15}\, (3-2\alpha)^{\ft32}}\,,
\qquad
\fft{Q}{\wtd Q} = \fft{\sqrt2\, \alpha^2}{\sqrt3\, (3-2\alpha)^{\ft32}}\,,
\label{B2ratios}
\eea
where we have defined $\alpha= \dfft{c_2^2}{c_1^2}$ with $c_1$ and $c_2$ being the two parameters employed in \cite{luwuwuzh} (they were introduced there as integration constants).  The dilaton coupling constants for this $B_2$ Toda case are \cite{luwuwuzh}
\bea
a= 3\,,\qquad b=-2\,.\label{B2ab}
\eea
Comparing equations (\ref{B2ratios}) and (\ref{B2ab}) with our expressions (\ref{MSigsols}), it can be seen that in this case $x$ is related to $\alpha$ by 
\bea
e^{5x} = \fft{\sqrt2\, \alpha^2}{\sqrt3\, (3-2\alpha)^{\ft32}}\,,
\label{B2x}
\eea
and the function $h(x)$, viewed now as a function of $\alpha$, is given by
\bea
h = \fft{k\, (9-3\alpha - \alpha^2)}{\alpha^{\ft45}\, (3-2\alpha)^{\ft{9}{10}}}
\,,\qquad \hbox{where}\quad k= 2^{-\ft15}\, 3^{-\ft3{10}}\, 5^{-\ft12}\,.
\label{B2h}
\eea
Although we cannot usefully invert eqn (\ref{B2x}) to give $\alpha$ as a function of $x$, it can nevertheless be seen by re-expressing $h'(x)$ as $h'(x) = \dfft{dh}{d\alpha}\, \dfft{d\alpha}{dx}$ that eqns (\ref{B2x}) and (\ref{B2h}) do indeed provide an exact solution of our mass equation (\ref{diffeq}), provided that eqns
(\ref{B2ab}) hold.

It is worth noting that one can eliminate $\alpha$ between the
two equations (\ref{B2x}) and (\ref{B2h}), giving
\crampest
\bea
100 e^{8x}\, h^4 - 20\sqrt{10}\, e^{11x}\, h^3 - 40 e^{4x}\, h^2 + 
2\sqrt{10}\, e^{7x}\, (e^{10x}-14)\, h + 4-22 e^{10x} -e^{20x}=0\,.
\label{B2hx}
\eea
\uncramp
Thus one cannot usefully give an explicit expression for $h(x)$
as a function of $x$ in this case, although one can use eqn
(\ref{B2hx}) in order to verify that eqn (\ref{diffeq}) is satisfied.

Using the expressions for the mass function $h(x)$ given in \eqref{B2h} above one can calculate the long-range force between two non-identical extremal black holes using \eqref{d3int}. This force between 
can be seen to be
\begin{align}
\frac{F_{12}}{{\wtd Q}_1\,{\wtd Q}_2} &=  \frac{{S_{12}}(\alpha_1,\,\alpha_2)}{(3-2\,\alpha_1)^\ft32\,(3-2\alpha_2)^\ft32} \,, \quad \text{where}\,, \nn\\
 S_{12}(\alpha_1,\,\alpha_2) &= 27\,(\alpha_1 + \,\alpha_2 -1) - 3\,(\alpha_1^2 + 10\,\alpha_1\,\alpha_2 + \alpha_2^2) + 4\,\alpha_1\,\alpha_2\,(\alpha_1+\alpha_2) \nn\\
 &\quad  + (3-2\,\alpha_1)^\ft32\,(3-2\alpha_2)^\ft32\,. 
\end{align}
Here $\alpha_1$ and $\alpha_2$ refer to the parameter $\alpha$ for each black hole. After a change of variable given by\footnote{Although the expressions (\ref{B2ratios}) would appear
to allow any parameter value $\alpha$ obeying $\alpha\le \ft32$, it can be seen, by looking at the
explicit form of the extremal black hole metrics, that there will be naked singularities
outside the horizon if $\alpha$ is negative.  Thus we must restrict the parameter range
to $0\le \alpha\le \ft32$.  This is discussed in more detail below.}
\bea
\alpha_i = \ft32\,\sin^2 t_i \,, \quad t_i \in [0,\, \fft{\pi}2]\,,
\eea
the force can be brought to the following form:
\bea
\frac{F_{12}}{{\wtd Q}_1\,{\wtd Q}_2} = \frac{(\cos t_1 - \cos t_2)^2}{8\cos t_1^3\,\cos t_2^3}\,\Big[\sin^2 (t_1-t_2) + \sin^2 (t_1+t_2) + 4\cos t_1\cos t_2 \Big]\, \geq 0\,. 
\eea
Note that this implies that the force is always repulsive. This is in perfect agreement with the general trend of mutual force in four dimensions \eqref{Ftrend4D}, since for $B_2$ Toda black holes $a=3,\,b=-2$, hence $b < -1/a$.

Note that, just as in $A_2$ black holes, the mass function for $B_2$ Toda solutions can be straightforwardly extended to higher dimensions using \eqref{rescale}, and the non-negativity of the mutual force will continue to hold in higher dimensions, as the mutual force expression, \eqref{Mintdunequal2}, remains identical to the one in four dimension.
\newpage

\bigskip
{\bf Remark on the paramater range for $\alpha$}:
\medskip

A priori, there is no obvious problem in the formulae (\ref{B2ratios}) with allowing $\alpha$ to be negative.  However, if we look at the metric functions $H_1$ and $H_2$ for the extremal $B_2$ black hole, as given in eqns (60) of \cite{luwuwuzh}, it can be seen that, introducing a radial coordinate $z=c_1\, \rho=
c_1\, r^{-1}$, we have
\bea
H_1 = \fft{(1+\alpha\, z )(3-2\alpha + 3\alpha\, z + 3\alpha^2\, z^2 + 
   \alpha^3\, z^3)}{3-2\alpha}\,,
\eea
and therefore if $\alpha$ were negative the function $H_1$ would have a zero at
$z=-\alpha$, which would be greater than zero.  Since this is located outside the extremal horizon, the solution would have a naked singularity in this case. By contrast, the
metric function $H_2$ in eqns (60) of \cite{luwuwuzh} has no real zeros outside the
horizon, regardless of the sign of $\alpha$. In summary, therefore, in order to have regular extremal black holes we must require $\alpha$ to be positive.

\bigskip
\noindent\underline{{\bf $G_2$ Toda black hole}}:
\medskip

From the results in eqns (71) of \cite{luwuwuzh}, in four dimensions the mass and electric charges for the extremal  $G_2$ Toda black hole will be such that, in our normalisations, 
\bea
\fft{M}{\wtd Q} = 
\fft{\sqrt3\, (1- q^2 -9 q^4 + q^6 )}{
   2\sqrt 7\, (1- 4 q^2 + q^4)^{\ft32}}\,,
\qquad
\fft{Q}{\wtd Q} = \fft{2\sqrt{6}\, q^5}{(1- 4 q^2 + q^4)^{\ft32}}
\,,
\label{G2ratios}
\eea
where we have defined $q= \dfft{c_2^3}{\sqrt5\, c_1^3}$, with $c_1$ and $c_2$ being the two parameters introduced in \cite{luwuwuzh}.  
Note that if we require $M/\tilde{Q}$ and $Q/\tilde{Q}$ to be real, and in addition
$Q/\tilde{Q}$ to be positive, we need 
\begin{equation}
0<q<(2-\sqrt{3})^{1/2} \quad \text{or} \quad q>(2+\sqrt{3})^{1/2} \,.\label{Qrealpos}
\end{equation}
The dilaton coupling constants for this $G_2$ Toda case are \cite{luwuwuzh}
\bea
a= 3\sqrt3\,,\qquad b=-\fft{5}{\sqrt3}\,.\label{G2ab}
\eea

It can in fact be seen that only the first of the two ranges for the parameter $q$ given in eqns
(\ref{Qrealpos}) corresponds to regular black holes; all the solutions for the second parameter range correspond to spacetimes with naked singularities outside the horizon. Since the 
demonstration of this fact is a little involved, we relegate the proof to subsection 
\ref{qrange} below.  From now on, then, we shall take the parameter $q$ to lie just in the
first of the two ranges in eqns (\ref{Qrealpos}), that is,
\bea
0\le q\le q_{\rm max}\,,\qquad \hbox{where}\qquad q_{\rm max} = \sqrt{2-\sqrt3}\,.
\label{G2qrange}
\eea

From eqns (\ref{MSigsols}) we can read off that for this $G_2$ extremal black hole we have\footnote{Another way to see that the parameter $q$ should be positive is from the relation above between $x$ and $q$.}
\bea
e^{\ft{14}{\sqrt3}\, x} &=& \fft{2\sqrt{6}\, q^5}{(1-4q^2+q^4)^{\ft32}}\,,\nn\\
h &=& \fft{k\, (1-q^2-9q^4+q^6)}{
 q^{\ft{25}{14}}\, (1 - 4 q^2 + q^4)^{\ft{27}{28}}}\,,
 \qquad
 \hbox{where}\quad k= 2^{-\ft{43}{28}}\, 3^{\ft{9}{28}} \, 
 7^{-\ft12}\,.\label{xhG2}
 \eea
It can be verified that this does indeed provide an exact solution of the mass equation (\ref{diffeq}), provided that $a$ and $b$ are given by eqns (\ref{G2ab}).

Note that we can eliminate $q$ between the two equations (\ref{xhG2}), yielding a purely algebraic relation between $x$ and $h(x)$.  To simplify the expression we define $\hat h(x)$ and $w(x)$ by writing
\bea
h(x)= \fft1{2\sqrt7}\, e^{-\ft{5x}{\sqrt3}}\, \hat h(x)\,,\qquad w(x) = e^{\ft{28 x}{\sqrt3}}\,,
\eea
in terms of which we find
\bea
0&=& \hat h^{12} - 6\hat h^{10}\, (3+w) + 3\hat h^8\, (45-348 w + 5 w^2) 
-2 \hat h^6\, (270 + 20241 w -1449 w^2 + 10 w^3) \nn\\
&& + 3\hat h^4\, (405 -16722 w + 2061 w^2 -750 w^3 + 5 w^4) \nn\\
&& - 6\hat h^2\, (243 -82782 w + 425025 w^2 + 27837 w^3 - 12 w^4 + w^5)\nn\\
&& + (27+2601 w + 171 w^2 + w^3)^2\,.
\eea
It can also be verified directly from this relation that $h(x)$ obeys the mass equation (\ref{diffeq}) when $a$ and $b$ are given by eqns (\ref{G2ab}).

As already mentioned, the equivalent solutions to the mass equation in dimensions greater than four can be obtained straightforwardly from the results we have obtained above, simply by implementing the transformations given in eqns (\ref{rescale}) and (\ref{albe}).

In a similar fashion to our previous discussion for the $B_2$ Toda black holes, we can obtain an 
expression for 
the mutual force between unequal extremal $G_2$ Toda black holes by using eqn (\ref{d3int}). In this case, the resulting expression is somewhat involved, with the general force expression
being given by
\bea
F_{12}= \fft{\wtd Q_1\, \wtd Q_2\, S_{12}}{(1-4q_1^2 + q_1^4)^{3/2}\,
(1-4q_2^2 + q_2^4)^{3/2}}\,,
\eea
with 
\bea
S_{12} &=& -1 -q_1^6\, q_2^6 
+ (q_1^2+q_2^2)(6-q_1^4-q_2^4+ 46 q_1^2\, q_2^2) -
6 q_1^4\, q_2^4\, (q_1^2+q_2^2-4 q_1\, q_2) \nn\\
&&
+ 3q_1^2\, q_2^2\, (2q_1^4
+ 2q_2^4 - 21 q_1^2\, q_2^2) -3 (2q_1^4 + 2q_2^4 + 13q_1^2\, q_2^2) \nn\\
&& +
(1-4 q_1^2 + q_1^4)^{3/2} \,(1-4 q_2^2+q_2^4)^{3/2}\,.
\eea
The parameters $q_1$ and $q_2$ are constrained to lie in the intervals 
\bea
0\le q_1\le q_{\rm max}\,,\qquad 0\le q_2\le q_{\rm max}\,,\label{G2square}
\eea
where $q_{\rm max}$ is given in eqn (\ref{G2qrange}).

\begin{figure}[h!]
        \centering
        \includegraphics[width=0.6\linewidth]{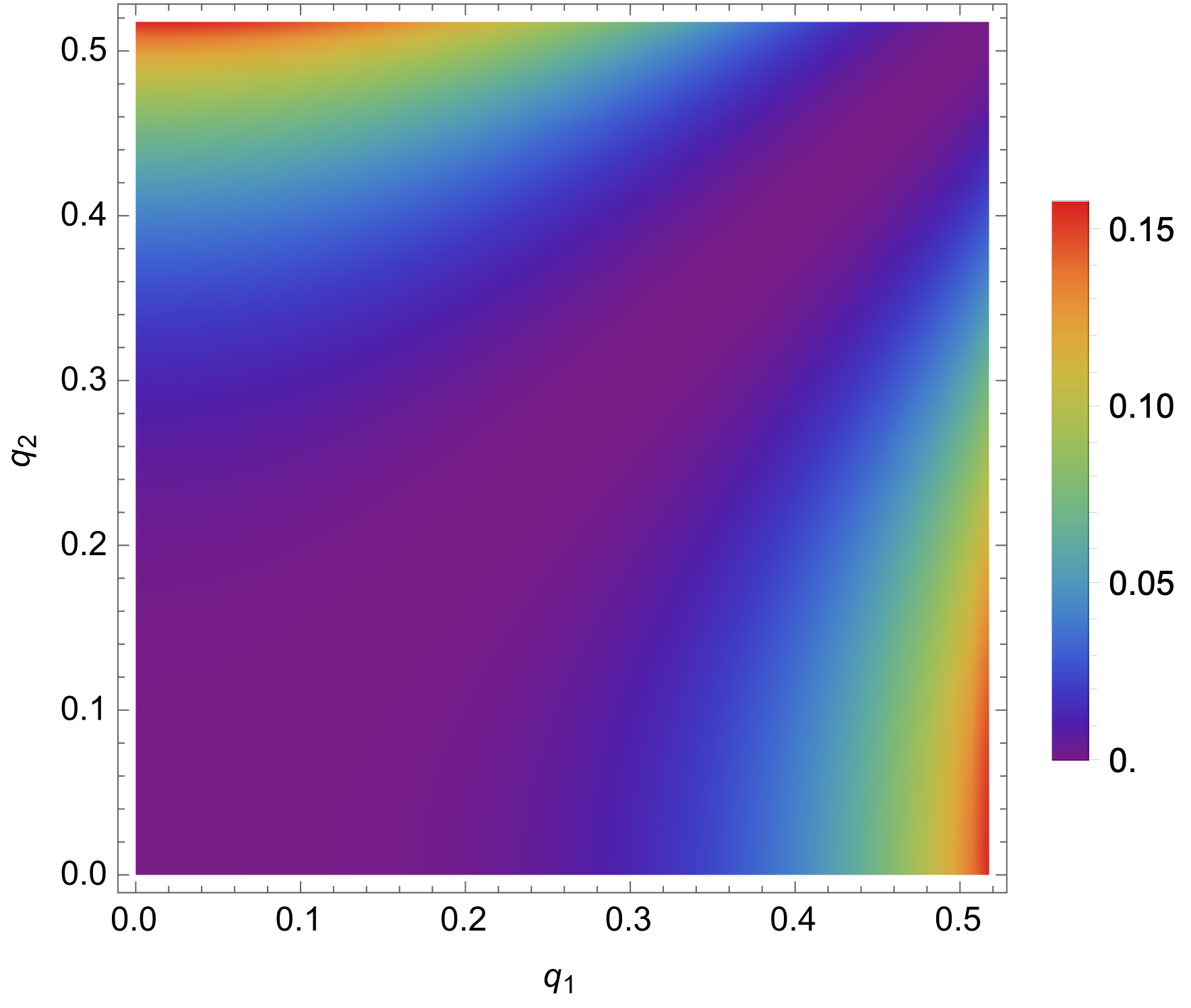}
        \caption{Density plot of $S_{12}(q_1,q_2)$ showing that $S_{12}\geq 0$ when $q_1,q_2$ lie within $(0,q_\text{max})$.}
        \label{G2plot}
    \end{figure}

According to the conjectured trend for the mutual force, for $G_2$ Toda black holes we should expect the force to be positive or repulsive, since $a = 3\sqrt{3},\, b = -5/\sqrt{3}$ and so $b < -1/a$. In fig \eqref{G2plot} below we verify by making a density plot of $S_{12}(q_1,q_2)$ for $q_1$ and $q_2$ in 
the square defined by the permitted ranges (\ref{G2square}) that the function $S_{12}$ is
indeed non-negative.

It vanishes on the diagonal line $q_2=q_1$, and as can be seen from the density plot,
it is in the region near to this line that the risk of becoming negative is the greatest.  It is instructive,
therefore, to look analytically at this region, which we can do by considering
\begin{equation}
    q_1= q+ \frac{\delta}{2}\, , \quad  q_2= q- \frac{\delta}{2}\, , \quad |\delta|<<1 \,.
\end{equation}
The long range force is proportional to\begin{equation}
    \frac{F_{12}}{\tilde{Q}_1 \tilde{Q}_2}= \frac{6 q^2 (1+9q^4-2 q^6)}{(1-4 q^2+q^4)^3} \,\delta^2 + \mathcal{O}(\delta^4)
\end{equation}
which is positive for the extremal black hole solutions, since $(1+9q^4-2q^6)>0$ for $0\le q\le q_{\rm max}$.

For higher dimensional extension, the mass function and the mutual force for $G_2$ solutions are immediately obtained from \eqref{rescale} and \eqref{Mintdunequal2}. Since the higher dimensional mutual force in terms of the rescaled quantities ${\hat h},\, {\hat x}$ is identical to the four dimensional one, the non-negativity of the force should continue to hold.

\subsection{Ruling out the $q>(2+\sqrt3)^{1/2}$ branch of extremal $G_2$ black holes
\label{qrange}}

The second branch of the $q$ ranges specified in eqns (\ref{Qrealpos}) can be ruled out on the grounds that one of the two metric functions $H_1$ and $H_2$, specifically $H_1$
(see \cite{luwuwuzh}), will have zeros outside the extremal horizon, and thus such black holes will have naked singularities.  Introducing a radial variable
$z =\sqrt5\, c_1\, \rho$, where $\rho=\dfft1{r}$, the metric functions $H_1$ and $H_2$
for the $G_2$ extremal black hole, given in eqns (70) of \cite{luwuwuzh}, can be written as
\bea
H_1 &=& 1 + (1-4q^2+q^4)^{-1}\, \Big[6q\, (1-3q^2)\, z +
 18 q^2\, (1-2q^2)\, z^2\nn\\
 &&\qquad + 36 q^3(1-q^2)\, z^3 + 18 q^4\, (3-q^2)\, z^4 + 
 18 q^5\, (17-q^2)\, z^5\nn\\
 &&\qquad +
  \fft{252}{5} q^6\, z^6 + \fft{144}{5}\, q^7\, z^7 + \fft{54}{5}\, q^8\, z^8 
  + \fft{12}{5}\, q^9\, z^9 + \fft{6}{25} q^{10}\, z^{10}\Big]\,,\nn\\
H_2 &=& 1+ \fft{1+q^2}{q} \, z + 3 z^2 + 4 q\, z^3 +
  3 q^2\, \, z^4 + \fft{6}{5}\, q^3\, z^5 + \fft1{5}\, q^4\, z^6\,.\label{G2H1H2}
\eea
Note that $z=\infty$ corresponds to the location of the extremal horizon, and $z=0$ corresponds to the
asymptotic region infinitely far from the black hole.
 
In fact, the metric function $H_1$ will always have zeros outside the extremal horizon when $q$ lies in the second branch, $q\ge (2+\sqrt3)^{1/2}$, in eqns (\ref{Qrealpos}).  
This can be shown by using Sturm's theorem \cite{sturm,sturmwiki} 
to determine the number of real zeros of $H_1$ in the interval
$0\le z\le \infty$.  First, we clear the denominator in the expression for
$H_1$ in eqns  (\ref{G2H1H2}), defining $f(z)= (1-4 q^2 + q^4)\, H_1(z)$, so
\bea
f(z)&=& (1-4q^2+q^4) + 6q\, (1-3q^2)\, z +
 18 q^2\, (1-2q^2)\, z^2\nn\\
 &&+ 36 q^3(1-q^2)\, z^3 + 18 q^4\, (3-q^2)\, z^4 + 
 18 q^5\, (17-q^2)\, z^5\nn\\
 && +
  \fft{252}{5} q^6\, z^6 + \fft{144}{5}\, q^7\, z^7 + \fft{54}{5}\, q^8\, z^8 
  + \fft{12}{5}\, q^9\, z^9 + \fft{6}{25} q^{10}\, z^{10}\,,\label{G2fdef}
\eea
Next, we construct the Sturm sequence for this polynomial, which in this case has length five, $\{P_0, P_1, P_2,P_3,P_4\}$, with\footnote{Generically, since $f(z)$ is of degree 10, the Sturm sequence could have been expected to have length 10. Evidently, the polynomial $f(z)$ must have some special features here.}
\bea
P_0 &=& f(z)\, \qquad P_1= f'(z)\,,\nn\\
P_2&=& -\fft{(1-5 q^2)(10-5 q^2 +45 q\, z + 90q^2\, z^2 + 90 q^3\, z^3 +
   +45 q^4\, z^4 +  9 q^5\, z^5)}{25}\nn\\
P_3 &=& -\fft{10 q\, (1-5q^2)\, (1+q\, z)^4}{3}\,,\nn\\
P_4&=& \fft{(1-5 q^2)^2}{25}\,.
\eea
Sturm's theorem states that in any interval $a\le z\le b$, if the number of sign changes as one works along the ordered list $\{P_0,P_1,P_2,P_3,P_4\}$ when $z=z_1$
is called $V(z_1)$, and if the number of sign changes when $z=z_2$ is called $V(z_2)$, then the number
of real roots in the interval $z_1\le z\le z_2$ is equal to $V(z_1)-V(z_2)$.  In our case, it
is easily seen that for the second branch in eqns (\ref{qrange}), 
where $q\ge (2+\sqrt3)^{1/2}$, the ordered
lists of signs at $z_1=0$ and $z_2=\infty$ are:
\bea
z=0:&&  \{+,-,-,+,+\}\,,\nn\\
z=\infty: && \{+,+,+,+,+\}\,.
\eea
Thus we have $V(0)=2$ and $V(\infty)=0$, and hence the function $f(z)$ has two real zeros in the interval $0\le z\le\infty$.  This shows that the $q\ge (2+\sqrt3)^{1/2}$ branch of the extremal $G_2$ black holes will all have naked singularities, and thus should be excluded.

By contrast, for the first branch in eqns (\ref{qrange}), where $0\le q\le (2-\sqrt3)^{1/2}$, the signs are either
\bea
z=0: && \{+,+,+,+,+\}\,,\nn\\\
z=\infty:&& \{+,+,+,+,+\}
\eea
when $\dfft{1}{\sqrt5} < q\le (2-\sqrt3)^{1/2}$, in which case $V(0)=V(\infty)=0$ and so there are no real roots in the interval $0\le z\le\infty$, or else the signs are
\bea
z=0: && \{+,+,-,-,+\}\,,\nn\\
z=\infty: && \{+,+,-,-,+\}\,,
\eea
when $0\le q\le \dfft1{\sqrt5}$, in which case $V(0)=V(\infty)=2$ and so again there are no real roots in the interval $0\le z\le\infty$.  There are never any zeros
for the metric function $H_2$ in the interval $0\le z\le\infty$.  

In summary, the branch of extremal $G_2$ black holes with $0\leq q \le (2-\sqrt3)^{1/2}$ are regular, in the sense of having no naked singularities outside the horizon.  By contrast, the branch with $q\ge (2+\sqrt3)^{1/2}$ all have naked singularities outside the horizon, and these should be rejected as valid black hole solutions.  Once this branch is rejected, all the remaining solutions with unequal charges experience a repulsive long range force, which is in accordance with the results of our analysis in the main body of the paper.

\subsection{A remark on algebraic relations between $M$, 
$Q$ and $\wtd Q$}

In each of the cases of the $A_2$, $B_2$ and $G_2$ extremal Toda black holes one can find a multinomial relating $M$, $Q$ and 
$\wtd Q$.  For example, for the $A_2$ extremal black hole 
one can rewrite eqn (\ref{A2mass}) as the multinomial
\bea
0&=& 64 M^6 - 48 M^4\, \big(Q^2+\wtd Q^2\big)
+ 12 M^2\,\big(Q^2+3Q\, \wtd Q+\wtd Q^2\big)
\big(Q^2-3Q\, \wtd Q+\wtd Q^2\big)\nn\\
&&-\big(Q^2+\wtd Q^2\big)^3\,.
\eea

In the case of $B_2$ we can eliminate $\alpha$ between the two equations in (\ref{B2ratios}).  The resulting eighth-order multinomial in $M$, $Q$ and $\wtd Q$ actually factorises over the rationals as a product of two quartic multinomials if one defines $Q=\sqrt{10}\, Q'$ and
$\wtd Q= \sqrt{10}\, \wtd Q'$, in the form
\bea
P_+(M,Q',\wtd Q')\, P_-(M,Q',\wtd Q') =0\,,
\eea
where
\bea
P_+(M,Q',\wtd Q') 
&=& M^4 - 2 M^3\, Q' - 4 M^2\, {\wtd Q'}{}^2 + 2M\, Q'\, 
({Q'}{}^2 - 14 {\wtd Q'}{}^2) \nn\\
&&-2{\wtd Q'}{}^2\, (11{Q'}{}^2 -2{\wtd Q'}{}^2)\,,
\eea
and $P_-(M,Q',\wtd Q')=P_+(M,-Q',-\wtd Q')$.

For the $G_2$ case, if we eliminate $q$ between the two expressions in (\ref{G2ratios}) we find, after defining $(Q',\wtd Q')=(2\sqrt7\, Q, 2\sqrt7\, \wtd Q)$ that $M$, $Q'$ and $\wtd Q'$ are related by
\bea
0&=& M^{12} - 6 M^{10}\, \big(Q'{}^2 + 3\wtd Q'{}^2\big) 
 + 3 M^8\, 
\big(5 Q'{}^4 -348 Q'{}^2\, \wtd Q'{}^2 + 45 \wtd Q'{}^4\big) 
\nn\\
&&-2M^6\,\big(10 Q'{}^6 - 1449 Q'{}^4\, \wtd Q'{}^2 +
20241 Q'{}^2\, \wtd Q'{}^4 + 270 \wtd Q'{}^6\big)\nn\\
&& +3M^4\, \big(5 Q'{}^8 -750 Q'{}^6\, \wtd Q'{}^2 +
2061 Q'{}^4\, \wtd Q'{}^4 - 16722 Q'{}^2 \, \wtd Q'{}^6 +
  405 \wtd Q'{}^8\big)\nn\\
&& -6M^2\, \big(Q'{}^{10} - 12 Q'{}^8\, \wtd Q'{}^2 +
 27837 Q'{}^6\, \wtd Q'{}^4 + 425025 Q'{}^4\, \wtd Q'{}^6 -
  82782 Q'{}^2\, \wtd Q'{}^8 + 243 \wtd Q'{}^{10}\big) \nn\\
&&+ \big(Q'{}^6 + 171 Q'{}^4\, \wtd Q'{}^2 - 
2601 Q'{}^2\, \wtd Q'{}^4 -27 \wtd Q'{}^6\big)^2\,.
\eea


\begin{thebibliography}{99}

\bibitem{Arkani-Hamed:2006emk}
N.~Arkani-Hamed, L.~Motl, A.~Nicolis and C.~Vafa,
{\it The String landscape, black holes and gravity as the weakest force,}
JHEP \textbf{06}, 060 (2007)
doi:10.1088/1126-6708/2007/06/060
[arXiv:hep-th/0601001 [hep-th]].

\bibitem{Palti:2017elp}
E.~Palti,
{\it The Weak Gravity Conjecture and Scalar Fields,}
JHEP \textbf{08}, 034 (2017)
doi:10.1007/JHEP08(2017)034
[arXiv:1705.04328 [hep-th]].

\bibitem{Heidenreich:2019zkl}
B.~Heidenreich, M.~Reece and T.~Rudelius,
{\it ``Repulsive Forces and the Weak Gravity Conjecture,}
JHEP \textbf{10}, 055 (2019)
doi:10.1007/JHEP10(2019)055
[arXiv:1906.02206 [hep-th]].

\bibitem{cvegibpop}
M.~Cveti\v c, G.W.~Gibbons and C.N.~Pope,
{\it Super-Geometrodynamics}, 
JHEP \textbf{03}, 029 (2015),
doi:10.1007/JHEP03(2015)029,
[arXiv:1411.1084 [hep-th]].

\bibitem{brillind} D.R.~Brill and R.W.~Lindquist,
{\it Interaction energy in geometrostatics},
Phys. Rev. \textbf{131}, 471-476 (1963),
doi:10.1103/PhysRev.131.471.

\bibitem{vinort} U.K.~Beckering Vinckers and T.~Ortin,
{\it On the interactions and equilibrium between Einstein-Maxwell-dilaton 
black holes},
SciPost Phys. Core \textbf{8}, 044 (2025),
doi:10.21468/SciPostPhysCore.8.2.044,
[arXiv:2408.04621 [gr-qc]].

\bibitem{gibkalkol} G.W.~Gibbons, R.~Kallosh and B.~Kol,
{\it Moduli, scalar charges, and the first law of black hole 
thermodynamics},
Phys. Rev. Lett. \textbf{77}, 4992-4995 (1996),
doi:10.1103/PhysRevLett.77.4992,
[arXiv:hep-th/9607108 [hep-th]].

\bibitem{gibmae} G.W.~Gibbons and K.i.~Maeda,
{\it Black Holes and Membranes in Higher Dimensional Theories with Dilaton Fields},
Nucl. Phys. B \textbf{298}, 741-775 (1988),
doi:10.1016/0550-3213(88)90006-5.

\bibitem{lu2field} H.~Lu,
{\it Charged dilatonic AdS black holes and magnetic AdS$_{D-2} \times R^{2}$ vacua},
JHEP \textbf{09}, 112 (2013),
doi:10.1007/JHEP09(2013)112,
[arXiv:1306.2386 [hep-th]].

\bibitem{kanpop} S.K.~Kanumilli and C.N.~Pope,
''Super-geometrodynamics in higher dimensions,''
Class. Quant. Grav. \textbf{35}, no.21, 214001 (2018),
doi:10.1088/1361-6382/aae32b,
[arXiv:1807.00039 [gr-qc]].

\bibitem{eihulesc} M.~Eichmair, L-H.~Huang, D.A.~Lee and R.~Schoen,
{\it The spacetime positive mass theorem in dimensions less than eight},
J. Eur. Math. Soc. 018, 83 (2016),  
 doi.org/10.4171/JEMS/584,
[arXiv:1110.2087v2 [math.DG]]

\bibitem{jau} J.L Jauregui, 
''Lower semicontinuity of the ADM mass in dimensions two through seven,''
Pacific Journal of Mathematics, 301, 441 (2019),
doi.org/10.2140/pjm.2019.301.441.


\bibitem{popetran}
M.~Cvetic, M.~J.~Duff, P.~Hoxha, J.~T.~Liu, H.~Lu, J.~X.~Lu, R.~Martinez-Acosta, C.~N.~Pope, H.~Sati and T.~A.~Tran,
``Embedding AdS black holes in ten-dimensions and eleven-dimensions,''
Nucl. Phys. B \textbf{558}, 96-126 (1999)
doi:10.1016/S0550-3213(99)00419-8
[arXiv:hep-th/9903214 [hep-th]].

\bibitem{Cremonini:2022sxf}
S.~Cremonini, M.~Cvetic, C.~N.~Pope and A.~Saha,
``Long-range forces between nonidentical black holes with non-BPS extremal limits,''
Phys. Rev. D \textbf{106}, no.8, 086007 (2022)
doi:10.1103/PhysRevD.106.086007
[arXiv:2207.00609 [hep-th]].

\bibitem{emdpaper} S.~Cremonini, M.~Cveti\v c, C.N.~Pope and A.~Saha,
{\it Mass and force relations for Einstein-Maxwell-dilaton black holes}, 
Phys. Rev. D \textbf{107}, no.12, 126023 (2023),
doi:10.1103/PhysRevD.107.126023,
[arXiv:2304.04791 [hep-th]].

\bibitem{Cremonini:2024eog}
S.~Cremonini, M.~Cvetic, C.~N.~Pope and A.~Saha,
``Mass and force relations for extremal Einstein-Maxwell-dilaton-axion black holes,''
Phys. Rev. D \textbf{111}, no.6, 066008 (2025)
doi:10.1103/PhysRevD.111.066008
[arXiv:2412.12277 [hep-th]].

\bibitem{luwuwuzh} 
H.~L\"u, P.Y.~Wu, Z.H.~Wu and W.~Zhao,
{\it Exact Toda Black Holes of Rank-2 Lie Groups},
[arXiv:2604.14288 [hep-th]].

\bibitem{sturm} J.C.F. Sturm, {\it M\'emoire sur la r\'esolution des \'equations num\'eriques"}, Bulletin des Sciences de F\'erussac. 11: 419–425, (1829).

\bibitem{sturmwiki} \url{https://en.wikipedia.org/wiki/Sturm's_theorem}

\bibitem{Cremonini:2021upd}
S.~Cremonini, C.~R.~T.~Jones, J.~T.~Liu, B.~McPeak and Y.~Tang,
``Repulsive black holes and higher-derivatives,''
JHEP \textbf{03}, 013 (2022)
doi:10.1007/JHEP03(2022)013
[arXiv:2110.10178 [hep-th]].

\end{thebibliography}
\end{document}